\documentclass[prd,amsmath,amssymb,superscriptaddress,preprintnumbers,twocolumn,10pt]{revtex4-1}

\pdfoutput=1 

\usepackage{graphicx}
\usepackage{dcolumn}
\usepackage{bm}

\usepackage{amssymb}
\usepackage{latexsym}
\usepackage{booktabs}
\usepackage{amsmath}
\usepackage{multirow}
\usepackage{url}

\usepackage{float}
\usepackage[colorlinks=true, linkcolor=red, citecolor=blue]{hyperref}

\usepackage[normalem]{ulem}
\usepackage{color}
\usepackage{array}
\usepackage{enumerate}

\usepackage{lineno}
\usepackage{graphicx}
\usepackage{dcolumn}
\usepackage{bm}
\usepackage{amssymb}
\usepackage{latexsym}
\usepackage{booktabs}
\usepackage{amsmath}
\usepackage{multirow}
\usepackage{url}
\usepackage{footnote}
\usepackage{float}
\usepackage{acro}
\usepackage{hyperref}
\usepackage{cleveref}

\DeclareAcronym{PBH}{
  short = PBH ,
  long  = primordial black hole ,
  short-plural = s ,
}

\DeclareAcronym{CMB}{
  short = CMB ,
  long  = cosmic microwave background ,
  short-plural =  ,
}

\DeclareAcronym{LSS}{
  short = LSS ,
  long  = large-scale structures ,
  short-plural =  ,
}

\DeclareAcronym{IGM}{
  short = IGM ,
  long  = intergalactic medium ,
  short-plural =  ,
}

\DeclareAcronym{GW}{
  short = GW ,
  long  = gravitational wave ,
  short-plural = s ,
}

\DeclareAcronym{MCGs}{
  short = MCGs ,
  long  = molecular-cooling galaxies ,
  short-plural =  ,
}

\DeclareAcronym{ACGs}{
  short = ACGs ,
  long  = atom-cooling galaxies ,
  short-plural =  ,
}

\DeclareAcronym{EoR}{
  short = EoR ,
  long  = epoch of reionization ,
  short-plural =  ,
}

\DeclareAcronym{EDGES}{
  short = EDGES ,
  long  = Experiment To Detect The Global EoR Signature ,
  short-plural =  ,
}

\DeclareAcronym{INTEGRAL}{
  short = INTEGRAL ,
  long  = International Gamma-Ray Astrophysics Laboratory,
  short-plural =  ,
}

\DeclareAcronym{SARAS 3}{
  short = SARAS 3,
  long  = Shaped Antenna measurement of the background Radio Spectrum 3 ,
  short-plural =  ,
}

\DeclareAcronym{SKA}{
  short = SKA,
  long  = Square Kilometre Array ,
  short-plural =  ,
}

\DeclareAcronym{DM}{
   short = DM ,
   long  = dark matter ,
   short-plural =  ,
}
\DeclareAcronym{WIMP}{
   short = WIMP ,
   long  = weakly interacting massive particle ,
   short-plural = s ,
}

\DeclareAcronym{HEAO}{
   short = HEAO,
   long  = High Energy Astrophysical Observatory ,
   short-plural =  ,
}

\DeclareAcronym{COMPTEL}{
   short = COMPTEL ,
   long  = Imaging Compton Telescope ,
   short-plural =  ,
}

\DeclareAcronym{EGRET}{
   short = EGRET ,
   long  = Energetic Gamma-ray Experiment Telescope ,
   short-plural =  ,
}

\DeclareAcronym{VERITAS}{
   short = VERITAS ,
   long  = Very Energetic Radiation Imaging Telescope Array System ,
   short-plural = ,
}

\DeclareAcronym{H.E.S.S}{
   short = H.E.S.S,
   long  = High Energy Stereoscopic System ,
   short-plural = ,
}

\DeclareAcronym{MAGIC}{
   short = MAGIC,
   long  = Major Atmospheric Gamma Imaging Cherenkov telescopes ,
   short-plural =  ,
}

\DeclareAcronym{LIGO}{
   short = LIGO,
   long  = Laser Interferometer Gravitational-Wave Observatory ,
   short-plural =  ,
}

\DeclareAcronym{VIRGO}{
   short = VIRGO,
   long  = Virgo Gravitational Wave Interferometer  ,
   short-plural =  ,
}

\DeclareAcronym{MWA}{
   short = MWA,
   long  = Murchison Widefield Array ,
   short-plural =  ,
   }

\DeclareAcronym{LOFAR}{
   short = LOFAR,
   long  = Low Frequency Array ,
   short-plural =  ,
   }

\DeclareAcronym{HERA}{
   short = HERA,
   long  = Hydrogen Epoch of Reionization Array ,
   short-plural =  ,
   }

\DeclareAcronym{JWST}{
   short = JWST,
   long  = James Webb Space Telescope,
   short-plural =  ,
   }
\usepackage[english]{babel}

\usepackage{amsmath}
\usepackage{graphicx}

\begin{document}


\title{Prospects for probing dark matter particles and primordial black holes with the Square Kilometre Array using the 21 cm power spectrum at cosmic dawn}

\author{Meng-Lin Zhao}\email{zhaoml@stumail.neu.edu.cn}
\affiliation{Liaoning Key Laboratory of Cosmology and Astrophysics, College of Sciences, Northeastern University, Shenyang 110819, China}

\author{Yue Shao}\email{shaoyue0304@163.com}
\affiliation{Liaoning Key Laboratory of Cosmology and Astrophysics, College of Sciences, Northeastern University, Shenyang 110819, China}
\affiliation{Department of Physics, Liaoning Normal University, Dalian 116029, China}

\author{Sai Wang}\thanks{Corresponding author}\email{wangsai@hznu.edu.cn}
\affiliation{School of Physics, Hangzhou Normal University, Hangzhou 311121, China}

\author{Xin Zhang}\thanks{Corresponding author}\email{zhangxin@neu.edu.cn}
\affiliation{Liaoning Key Laboratory of Cosmology and Astrophysics, College of Sciences, Northeastern University, Shenyang 110819, China}
\affiliation{National Frontiers Science Center for Industrial Intelligence and Systems Optimization, Northeastern University, Shenyang 110819, China}
\affiliation{MOE Key Laboratory of Data Analytics and Optimization for Smart Industry, Northeastern University, Shenyang 110819, China}


\begin{abstract}
Probing the nature of dark matter (DM) remains an outstanding problem in modern cosmology. The 21 cm signal, as a sensitive tracer of neutral hydrogen during cosmic dawn, provides a unique means to investigate DM nature during this critical epoch. Annihilation and decay of DM particles, as well as Hawking radiation of primordial black holes (PBHs), can modify the thermal and ionization histories of the early universe, leaving distinctive imprints on the 21 cm power spectrum. Therefore, the redshifted 21 cm power spectrum serves as a powerful tool to investigate such DM processes. In this work, we systematically assess the potential of the upcoming Square Kilometre Array (SKA) to constrain DM and PBH parameters using the 21 cm power spectrum. Assuming $10,000$ hours of integration time, the SKA is projected to reach sensitivities of $\langle\sigma v\rangle \leq 10^{-28}\,{\rm cm}^{3}\,{\rm s}^{-1}$ and $\tau\geq 10^{28}\,{\rm seconds}$, for $10\,{\rm GeV}$ DM particles. It can also probe PBHs with masses of $10^{16}\,\mathrm{g}$ and abundances $f_{\mathrm{PBH}} \leq 10^{-6}$. These results indicate that the SKA could place constraints on DM annihilation, decay, and PBH Hawking radiation that are up to two to three orders of magnitude stronger than current limits. Furthermore, the SKA is expected to exceed existing bounds on sub-GeV DM and to probe Hawking radiation from PBHs with masses above $10^{17}\,{\rm g}$, which are otherwise inaccessible by conventional cosmological probes. Overall, the SKA holds great promise for advancing our understanding of both DM particles and PBHs, potentially offering new insights into the fundamental nature of DM.

\end{abstract}

\maketitle

\section{Introduction}\label{sec:intro}

\Ac{DM} remains a major unsolved problem in modern cosmology.
Astrophysical observations evidence that \ac{DM} comprises more than 80\% of the non-relativistic matter in the universe \cite{ParticleDataGroup:2024cfk,Rubakov:2019lyf}. 
Although numerous candidate models have been proposed to elucidate \ac{DM} properties, the fundamental nature of \ac{DM} remains unknown \cite{Bertone:2016nfn}.
{Particle \ac{DM} candidates that interact weakly with ordinary matter, including \acp{WIMP}, \ac{WIMP}-like particles, and sub-GeV particles, can annihilate or decay into standard model particles \cite{Liu:2019zez,Chen:2003gz}.}
These standard model particles can subsequently generate distinctive signatures, which can be probed by instruments such as radio telescopes, neutrino detectors, and cosmic-ray observatories \cite{Thorpe-Morgan:2024zcq,Gaskins:2016cha,Bertone:2004pz}. 
As non-particle \ac{DM} candidates, \acp{PBH} can generate standard model particles including photons, electron-positron pairs, and neutrinos via the process of Hawking radiation \cite{Hawking:1971ei}. 
Consequently, this process may also produce observable signatures \cite{Wang:2025lti,Barroso:2024cgg,Carr:2021bzv,Carr:2009jm,Zhang:2023tfv,Zhang:2023zmb,Jia:2025vqn}. 
However, no such evidence has been found.
These null detection results place stringent constraints on \ac{DM} parameters (e.g., annihilation cross section, decay lifetime, and \ac{PBH} mass and abundance), ruling out \ac{DM} models incompatible with current observational data.

The 21 cm signal, particularly its power spectrum, provides a promising probe for \ac{DM} \cite{Mohapatra:2025qpz,Shao:2024owi,Sun:2024ywb,Nishizawa:2024bnh,Zhao:2024jad,Shao:2023agv,Hiroshima:2021bxn,Saha:2021pqf,Cang:2021owu,Mena:2019nhm,Lopez-Honorez:2016sur}. 
\Ac{CMB} observations place constraints on \ac{DM} properties at recombination ($z\sim1100$) \cite{Xu:2024vdn,Capozzi:2023xie,Zhang:2023usm,Acharya:2020jbv,Chluba:2020oip,Planck:2018vyg,Clark:2016nst,Kawasaki:2015yya,Lopez-Honorez:2013cua,Slatyer:2009yq}, 
while cosmic rays and Lyman-$\alpha$ forest probe \ac{DM} at low redshifts ($z\leq6$) \cite{LHAASO:2025wgl,Saha:2024ies,Yang:2024vij,Koechler:2023ual,Calore:2022pks,Foster:2022nva,Bernal:2022swt,Cirelli:2020bpc,Wang:2020uvi,Boudaud:2018hqb,Boudaud:2018oya,HESS:2018kom,MAGIC:2017avy,VERITAS:2017tif,Boudaud:2016mos,Cohen:2016uyg,Carr:2016hva,Fermi-LAT:2015att,Massari:2015xea,HESS:2014zqa,Aleksic:2013xea,Diamanti:2013bia,Essig:2013goa,Cirelli:2009bb}. 
However, these approaches are ineffective at probing \ac{DM} during the cosmic dawn.
As a sensitive probe of neutral hydrogen, the 21 cm power spectrum captures the imprints of \ac{DM} during this epoch via \ac{DM}'s impact on the \ac{IGM}, thereby bridging this observational gap.
This occurs because \ac{DM} particle annihilation, decay, and \ac{PBH} Hawking radiation alter the thermal and ionization history of the \ac{IGM}, thereby modifying the spin temperature of neutral hydrogen and affecting the brightness temperature of the 21 cm signal.
The 21 cm power spectrum quantifies these brightness temperature fluctuations in three-dimensional space, thereby enabling the inference of \ac{DM} properties.

To date, no 21 cm power spectrum signal has been detected, while the next-generation radio telescopes are expected to probe this signal for the first time.
Existing radio telescopes, including the \ac{MWA} \cite{Kolopanis:2022mgk,Trott:2020szf}, the \ac{LOFAR} \cite{Mertens:2025pvk,2019MNRAS.488.4271G}, and the \ac{HERA} \cite{2022ApJ...925..221A}, have so far placed only upper limits on the 21 cm power spectrum.
For example, \ac{MWA} provided an upper limit of $(66.18\,\rm{mK})^2$ at redshift $z = 7.1$ with scale $k = 0.19\,h\,\rm{Mpc}^{-1}$ at the $95\%$ confidence level \cite{Kolopanis:2022mgk,Trott:2020szf}, while \ac{HERA} reported an upper limit on the 21 cm power spectrum of $(30.76\,\rm{mK})^2$ at the $95\%$ confidence level at $z = 7.9$ with $k = 0.19\,h\,\rm {Mpc}^{-1}$\cite{2022ApJ...925..221A}.
The strictest upper limit arises from \ac{LOFAR}, which is $(68.66\,\rm{mK})^{2}$ at $z= 10.1$ with $k = 0.076\,h\,\rm{Mpc}^{-1}$ at the $95\%$ confidence level \cite{Mertens:2025pvk,2019MNRAS.488.4271G}.
As the next-generation flagship radio telescope, the \ac{SKA}, with its high spectral resolution and wide field of view, is expected to enable the first precise measurements of the 21 cm power spectrum.
The \ac{SKA} construction has made progress, and its precursor array, \ac{SKA}-AA0.5, has successfully obtained its first scientific image \cite{Gagnon-Hartman:2025oxd}.
When operational, the \ac{SKA} will constrain \ac{DM} annihilation, decay, and \ac{PBH} Hawking radiation via their imprints on the 21 cm power spectrum, potentially providing unprecedented constraints on \ac{DM} properties, thus offering new insights into its fundamental nature.

In this work, we focus on the potential of the \ac{SKA} to investigate \ac{DM}.
We first simulate the 21 cm power spectrum that incorporates the effects of \ac{DM} particle annihilation, decay, and \ac{PBH} Hawking radiation.
We then utilize the Fisher information matrix to quantify the \ac{SKA}'s project sensitivity in constraining \ac{DM} parameters.
Finally, we propose and optimize observation strategies for the \ac{SKA} to probe \ac{DM}.
This work is expected to provide new avenues for studying the nature of \ac{DM}.

The structure of this paper is as follows.
Section \ref{sec:exoes} discusses the physical mechanisms of \ac{DM} energy injection into \ac{IGM}. Section \ref{sec:eeeps} shows the impact of energy injection on the 21 cm power spectrum.
Section \ref{sec:fmfor} provides an introduction to the Fisher information matrix analysis.
Section \ref{sec:skadp} quantifies the potential of the \ac{SKA} for constraining the \ac{DM} parameters. Section \ref{sec:sumdc} is the summary and discussion.

\section{Scenarios of exotic energy}\label{sec:exoes}

In this section, we describe the scenarios of \ac{DM}-induced exotic energy injected into \ac{IGM}. 
Throughout this work, $\rho_{\rm DM}$ encodes information on the distribution of \ac{DM}, indicating $\rho_{\rm DM}=\rho_{\rm DM}(z,\mathbf{x})$ with $z$ and $\mathbf{x}$, respectively, being the redshift and the comoving position.

\subsection{Annihilation and decay of DM particles}\label{sec:addmp}

The exotic energy can be injected into \ac{IGM} due to the annihilation and decay of \ac{DM} particles, denoted by $\chi$. 
In this work, we consider the annihilation channels of $\chi\chi\rightarrow\gamma\gamma$, $\chi\chi\rightarrow e^{+}e^{-}$, $\chi\chi\rightarrow b\bar{b}$, and the decay channels of $\chi\rightarrow\gamma\gamma$, $\chi\rightarrow e^{+}e^{-}$, $\chi\rightarrow b\bar{b}$.
For a given channel, the primary particles can generate the secondary particles due to the hadronization process, which is simulated using \texttt{PYTHIA} \cite{Bierlich:2022pfr} and \texttt{PPPC4DMID} \cite{Cirelli:2010xx}. 
In the following, we focus on photons and electron-positron pairs, which are either primary or secondary or both, since the exotic energy is deposited into the \ac{IGM} primarily via them \cite{Liu:2019bbm,Slatyer:2015kla,Slatyer:2015jla,Slatyer:2012yq}.

{In this work, we focus on the $s$-wave annihilation, which is characterized by zero relative orbital angular momentum, resulting in an approximately constant thermally-averaged annihilation cross-section. For this process, the energy injection rate per unit volume is given by}
\begin{eqnarray}
    \left(\frac{{\rm d}E}{{\rm d}V{\rm d}t}\right)_{\rm inj} = f_{\rm ann}^{2} \rho^{2}_{\rm DM} c^2 \frac{\langle \sigma v \rangle}{m_{\chi}}\ , \label{eq:darkmatterani}
\end{eqnarray}
where $f_{\rm ann}$ and $m_{\chi}$ stand for the fraction and mass of \ac{DM} particles that can annihilate, $\rho_{\rm DM}$ the energy density of \ac{DM}, $c$ the speed of light, and $\langle \sigma v \rangle$ the thermally averaged annihilation cross section for the given annihilation channel.

For the decay of \ac{DM} particles, we have the energy of photons and electron-positron pairs per unit volume per unit temporal interval as 
\begin{eqnarray}
    \left(\frac{{\rm d}E}{{\rm d}V{\rm d}t}\right)_{\rm inj} =f_{\rm dec} \rho_{\rm DM} c^2 \frac{1}{\tau}\ , \label{eq:darkmatterdecay} 
\end{eqnarray}
where $f_{\rm dec}$ stands for the fraction of \ac{DM} particles that can decay, and $\tau$ the lifetime of \ac{DM} particles for the given decay channel.

Throughout this work, we take $f_{\rm ann}=f_{\rm dec}=1$ for simplicity, but can quickly recover them if necessary.

\subsection{Hawking radiation of PBHs}\label{sec:hrpbh}

The exotic energy can also be injected into the \ac{IGM} via the Hawking radiation of \acp{PBH} \cite{Hawking:1971ei}. 
In this work, we focus on \acp{PBH} within the mass regime of $\sim10^{15}-10^{18}$\,g.
{This mass range is of particular interest because \acp{PBH} within it could constitute a significant fraction of \ac{DM} while evading existing constraints.
\acp{PBH} in this mass range have not yet evaporated, and their Hawking radiation is potentially detectable, contributing to the exotic energy injection into the \ac{IGM}.}
In this work, we consider the emission products in the form of photons and electron-positron pairs, since the exotic energy is deposited into the \ac{IGM} primarily via these particles \cite{Cang:2021owu,Saha:2021pqf,Mena:2019nhm}. 

We have their energy per unit volume per unit temporal interval as 
\begin{eqnarray}
    \left(\frac{{\rm d}E}{{\rm d}V{\rm d}t}\right)_{\rm inj} &&= \int_{0}^{5\,\rm GeV} \frac{{\rm d}^{2} N}{{\rm d} E {\rm d} t} \Big|_{\rm \gamma} n_{\rm PBH} E {\rm d} E \nonumber \\  && + \int_{m_{\rm e}c^{2}}^{5\,\rm GeV} \frac{{\rm d}^{2} N}{{\rm d} E {\rm d} t}\Big|_{\rm e} n_{\rm PBH} (E - m_{\rm e} c^{2}) {\rm d} E\ , \label{eq:Einjtotal}
\end{eqnarray}
where $\mathrm{d}^2 N/(\mathrm{d}E \mathrm{d}t)$ stands for the particle spectrum given by \texttt{BlackHawk} \cite{Auffinger:2020ztk}, $m_{e}$ the electron mass, and $n_{\rm PBH}$ the number density of \acp{PBH}. 
Here, $n_{\rm PBH}$ is related to the \ac{PBH} abundance, denoted by $f_{\rm PBH}$, as 
\begin{eqnarray}
    n_{\rm PBH} = \frac{f_{\rm PBH} \rho_{\rm DM} }{M_{\rm PBH}}\ , \label{eq:npbh}
\end{eqnarray}
where $M_{\rm PBH}$ stands for the \ac{PBH} mass. 
It should be noted that for simplicity we have assumed a monochromatic mass function of \acp{PBH}, though the extended ones can be incorporated if necessary.

\section{Imprints on the 21 cm power spectrum}\label{sec:eeeps}

In this section, following the conventions of Refs.~\cite{Facchinetti:2023slb,Pritchard:2011xb,Furlanetto:2006jb}, we demonstrate imprints of the injected exotic energy on the 21 cm brightness-temperature fluctuations at cosmic dawn. 
We simulate the 21 cm signal using a modified version of \texttt{21cmFAST} \cite{Mesinger:2010ne}.
{In this work, we modify the equations within \texttt{21cmFAST} for the gas temperature $T_{\rm K}$, the ionization fraction $x_{\rm e}$, and the Lyman-$\alpha$ coupling efficiency $x_{\alpha}$ to incorporate the effects of exotic energy injection from \ac{DM}. The modifications, detailed in Section~\ref{subsec:impofexo}, account for the heating, ionization, and Lyman-$\alpha$ flux induced by exotic injection. We employ the \texttt{darkhistory} code \cite{Liu:2019bbm} to compute the coefficiencies for exotic energy deposition through heating, ionization, and Lyman-$\alpha$ scattering, namely $F_{\rm heat}$, $F_{\rm HI}$, $F_{\rm He}$, and $F_{\rm exc}$, as described in Section~\ref{subsec:impofexo}.}

\subsection{21 cm power spectrum}

The differential brightness temperature of the 21 cm signal evaluated at the observer is defined by 
\begin{eqnarray}
    {\delta T_{b}}(\nu,\mathbf{x}) &&\simeq 23\, x_{\rm HI}(z,\mathbf{x}) \left(\frac{0.15}{\Omega_{\rm m}}\right)^{\frac{1}{2}} \left(\frac{\Omega_{\rm b} h^{2}}{0.02}\right) \left(\frac{1+z}{10}\right)^{\frac{1}{2}} \nonumber \\ && \times \left[1-\frac{T_{\rm CMB}(z)}{T_{\rm S}(z,\mathbf{x})}\right]~\rm{mK}\ , \label{eq:T21}
\end{eqnarray}
where $\nu=\nu_{21}/(1+z)$ is the redshifted frequency of 21 cm photons, $x_{\rm HI}$ the neutral fraction of hydrogen, $T_{\rm CMB}$ the temperature of the \ac{CMB}, $T_{\rm S}$ the spin temperature of neutral hydrogen,  $\Omega_{\rm m}$ (or $\Omega_{\rm b}$) the present-day energy-density fraction of non-relativistic (or baryonic) matter, and $h$ the dimensionless Hubble constant. 
Here, $T_{\rm S}$ is explicitly given by 
\begin{eqnarray}
    T_{\rm S}^{-1} = \frac{T_{\rm CMB}^{-1} + x_{\alpha}T_{\alpha}^{-1} + x_{\rm c}T_{\rm K}^{-1}}{1 + x_{\alpha} + x_{\rm c}}\ , \label{eq:TS}
\end{eqnarray}
where $T_{\alpha}$ stands for the color temperature of Lyman-$\alpha$ photons, $T_{\rm K}$ the kinetic temperature of the \ac{IGM} gas, $x_{\alpha}$ the Lyman-$\alpha$ coupling coefficient, and $x_{\rm c}$ the collisional coupling coefficient. 
Due to resonant scattering, $T_{\alpha}$ is tightly coupled to $T_{\rm K}$, namely $T_{\alpha}\simeq T_{\rm K}$.

The observable is defined as follows. 
We first define the fractional perturbation to the differential 21 cm brightness temperature as 
\begin{eqnarray}
    \delta_{21}(\nu,\mathbf{x}) = \frac{\delta T_{b}(\nu,\mathbf{x})}{\overline{\delta T_{b}}(\nu)} - 1 \ ,
\end{eqnarray}
where $\overline{\delta T_{b}}$ stands for the spatial average of $\delta T_{b}$. 
We further define the dimensionless power spectrum for $\delta_{21}$ as 
\begin{eqnarray}
    \langle \tilde{\delta}_{21}(z,\mathbf{k}) \tilde{\delta}_{21}^{\ast}(z,\mathbf{k}') \rangle = (2\pi)^{3}\delta(\mathbf{k}-\mathbf{k}')\frac{2\pi^{2}}{k^{3}}\Delta_{21}^{2}(z,k)\ ,
\end{eqnarray}
where $\tilde{\delta}_{21}(z,\mathbf{k})$ is the Fourier mode of $\delta_{21}(\nu,\mathbf{x})$ with $\nu$ being replaced with $z=\nu_{21}/\nu-1$ and $\mathbf{k}$ (or $k$) being the comoving wavevector (or wavenumber), $\langle ... \rangle$ the ensemble average, and $\delta(\mathbf{k}-\mathbf{k}')$ the Dirac delta function. 
Following the conventions of Ref.~\cite{Facchinetti:2023slb}, we define the 21 cm power spectrum in units of temperature as 
$\overline{\delta T_{b}}^{2}(z)\Delta_{\rm 21}^{2}(z,k)$, 
in which we have replaced $\nu$ with $z$ once again. 
In the following, the above observable would be frequently referred to.

\subsection{Imprints of the exotic energy}
\label{subsec:impofexo}
Once injected into \ac{IGM}, the exotic energy can be deposited into \ac{IGM}, thus altering the thermal and ionization histories of \ac{IGM}. 
It can also contribute to the Lyman-$\alpha$ flux, altering the Lyman-$\alpha$ coupling coefficient. 
Therefore, we expect the injected exotic energy to leave imprints on the 21 cm power spectrum.

Due to energy deposition processes, the exotic energy can heat and ionize the \ac{IGM} gas. 
The heating and ionizing rates per baryon, respectively, are given by 
\begin{eqnarray}
    \epsilon_{\rm exo,\rm heat} &=& F_{\rm heat}(z) \frac{1}{n_{\rm b}} \left(\frac{{\rm d} E}{{\rm d} V {\rm d} t}\right)_{\rm inj}\ , \label{eq:epsilonheat} \\
    \Lambda_{\rm exo,\rm ion} &=& F_{\rm HI}(z) \frac{1}{n_{\rm b}} \frac{n_{\rm H}}{n_{\rm b}} \frac{1}{E^{\rm HI}_{\rm ion}}  \left(\frac{{\rm d} E}{{\rm d} V {\rm d} t}\right)_{\rm inj} \nonumber \\&+& F_{\rm He}(z)   \frac{1}{n_{\rm b}} \frac{n_{\rm He}}{n_{\rm b}} \frac{1}{E^{\rm He}_{\rm ion}} \left(\frac{{\rm d} E}{{\rm d} V {\rm d} t}\right)_{\rm inj}\ . \label{eq:lamdaexo}
\end{eqnarray}
Here, we adopt the delayed energy deposition model that is integrated in \texttt{Darkhistory}.
$F_{\rm heat}$, $F_{\rm HI}$, and $F_{\rm He}$ represent the energy deposition efficiencies via the processes of \ac{IGM} heating, hydrogen ionization, and helium ionization, respectively.
These deposition efficientcies are estimated by the \texttt{darkhistory}. 
$n_{\rm b}$, $n_{\rm H}$, and $n_{\rm He}$ denote the number densities of baryons, hydrogen, and helium, respectively. 
$E^{\rm HI}_{\rm ion}$ and $E^{\rm He}_{\rm ion}$ are the ionization energies of hydrogen and helium, respectively.

Considering both astrophysical processes and exotic energy, we have the heating and ionizing equations of the \ac{IGM} gas as 
\begin{eqnarray}
    \frac{{\rm d} T_{\rm K}(z,\mathbf{x})}{{\rm d} z} &=&\frac{2}{3 k_{\rm B} (1+x_{\rm e})} \frac{{\rm d} t}{{\rm d} z} (\epsilon_{\rm exo,\rm heat}+\epsilon_{\rm X,\rm heat}+\epsilon_{\rm IC,\rm heat}) \nonumber \\ &+& \frac{2 T_{\rm K}}{3 n_{\rm b}} \frac{{\rm d} n_{\rm b}}{{\rm d} z} - \frac{T_{\rm K}}{1 + x_{\rm e}} \frac{{\rm d} x_{\rm e}}{{\rm d} z}\ , \label{eq:TK} \\
    \frac{{\rm d} x_{\rm e}(z,\mathbf{x})}{{\rm d} z} &=& \frac{{\rm d} t}{{\rm d} z} (\Lambda_{\rm exo,\rm ion} + \Lambda_{\rm X,\rm ion} - \alpha_{\rm A} C x^{2}_{\rm e} n_{\rm H})\ . \label{eq:xe}
\end{eqnarray}
Here, $T_{\rm K}$ and $x_{\rm e}=1-x_{\rm HI}$, respectively, represent the kinetic temperature and ionization fraction. 
$k_{\rm B}$ is the Boltzmann constant. 
$t$ is the cosmic time. 
$\epsilon_{\rm X,\rm heat}$ and $\epsilon_{\rm IC,\rm heat}$, respectively, stand for the heating rates per baryon due to astrophysical X-rays and inverse-Compton scattering. 
$\Lambda_{\rm X,\rm ion}$ is the ionization rate due to astrophysical X-rays. 
$\alpha_{\rm A}$ denotes the case-A recombination coefficient. 
$C$ is the clumping factor. 
We have modified the corresponding equations in \texttt{21cmFAST}. 

Due to the deposition, the exotic energy can also have a contribution to the Lyman-$\alpha$ flux, i.e., 
\begin{eqnarray}
    J_{\alpha,\rm exo} = F_{\rm exc}(z) \frac{1}{n_{\rm b}} \frac{c n_{\rm b}}{4 \pi} \frac{1}{E_{\alpha}} \frac{1}{H(z) \nu_{\alpha}}\left(\frac{{\rm d} E}{{\rm d} V {\rm d} t}\right)_{\rm inj}\ . \label{eq:Jexo}
\end{eqnarray}
Here, $F_{\rm exc}$ represents the energy deposition efficiency via the process of hydrogen excitation. 
It is also given by \texttt{darkhistory}. 
$E_{\alpha}$ and $\nu_{\alpha}$, respectively, denote the energy and frequency of Lyman-$\alpha$ photons. 
$H(z)$ is the Hubble parameter at $z$. 

\begin{figure*}
    \centering
      \includegraphics[width=1.0\linewidth]{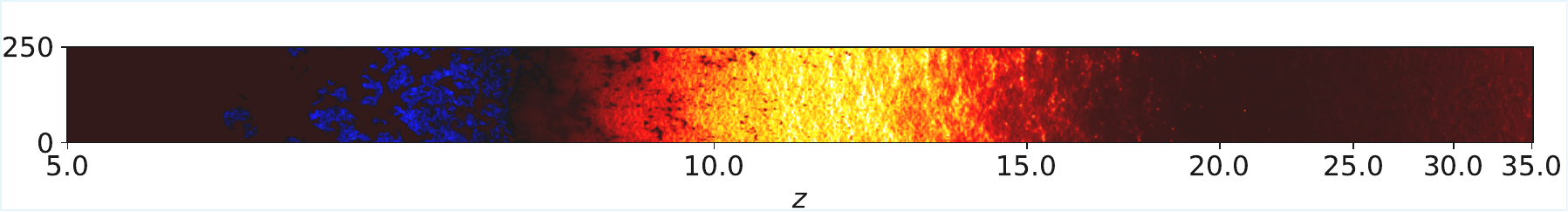}
    \includegraphics[width=1.0\linewidth]{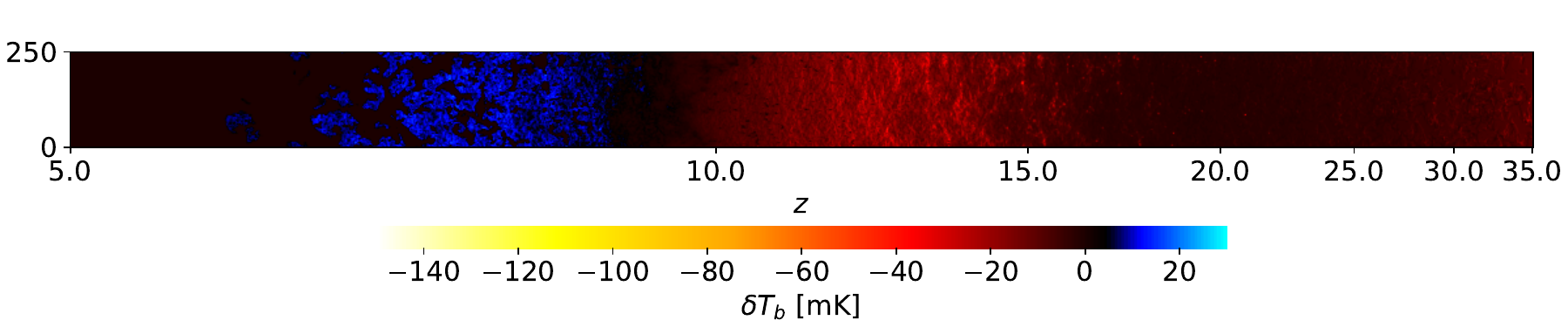}
    \caption{Lightcone slices of the differential brightness temperature in our $(250~{\rm Mpc})^{3}$ large
 simulation box. The fiducial model are shown on the upper panel. The bottom panel show the slice with \ac{DM} particle annihilation through $\chi\rightarrow e^{+}e^{-}$ channel with $ m_{\chi} = 10 ~\rm{GeV}$ and $\tau = 10^{27}\, s$.}
    \label{fig:lightcone_slice}
\end{figure*}

Taking into account Eq.~(\ref{eq:Jexo}), we modify the Lyman-$\alpha$ coupling coefficient to 
\begin{eqnarray}
    x_{\alpha} = \frac{1.7 \times 10^{11}}{1 + z} S_{\alpha} (J_{\alpha,\rm exo} + J_{\alpha,\rm X} + J_{\alpha,\rm \star})\ . \label{eq:xa}
\end{eqnarray}
Here, $S_{\alpha}$ denotes a quantum-mechanical correction factor. $J_{\alpha,\rm X}$ and $J_{\alpha,\rm \star}$, respectively, stand for the Lyman-$\alpha$ fluxes contributed by astrophysical X-rays and stellar emissions. 
We have also modified the corresponding equations in \texttt{21cmFAST}. 
{In Fig.~\ref{fig:lightcone_slice}, we show the lightcone slices of our fiducial model and of a model with \ac{DM} particle decay for comparison.}

\section{Fisher information matrix}\label{sec:fmfor}

To quantify the sensitivity of the \ac{SKA} in probing \ac{DM} particles and \acp{PBH}, we employ the Fisher information matrix.
Assuming Gaussian posterior distributions for relevant parameters, the Fisher matrix for the 21 cm power spectrum is given by \cite{Facchinetti:2023slb}
\begin{eqnarray}
    F_{ij} &=& \sum_{l}^{N_z}\sum_{m}^{N_k} \frac{1}{\sigma_{\rm tot}^2(z_{l},k_{m})}\frac{\partial [\overline{\delta T_{b}}^{2}(z_{l})\Delta_{21}^2(z_{l},k_{m} )]} {\partial \theta_i} \nonumber \\&\times& \frac{\partial
    [\overline{\delta T_{b}}^{2} (z_{l})\Delta_{21}^2(z_{l},k_{m} )]} {\partial \theta_j}\ . 
    \label{eq:FisherMatrix_power}
\end{eqnarray}
{In this work, we discretize the 21 cm power spectrum into $N_k \times N_z$ independent bins, following Ref.~\cite{Mason:2022obt}.}
{$N_k$ and $N_z$ represent the numbers of linearly discretized bins in $k$ and $z$, respectively.}
{The $k$ range is from 0.2 to 0.9 $\rm Mpc^{-1}$, and the $z$ range is from 6 to 20.}
$\sigma_{\rm tot}^2(z_{l},k_{m})$ is the total noise for the 21 cm power spectrum in the redshift bin $z_{l}$ and wavenumber bin $k_{m}$.
$\theta_{i}$ and $\theta_{j}$ represent the $i$-th and $j$-th parameter in the parameter set.

The total noise on the 21 cm power spectrum measurement arises from three key sources \cite{Pober:2012zz}
\begin{eqnarray}
    \sigma_{\rm tot}^2 \equiv \left[0.2 \overline{\delta T_{b}}^{2}\Delta_{21}^2\right]^2 + \sigma_{\rm poisson}^2 + \sigma_{\rm ins}^{2}\ , \label{eq:noise}
\end{eqnarray}
where the first term represents a conservative $20\%$ theoretical uncertainty in modeling the 21 cm signal \cite{Zahn:2010yw}, the second term quantifies the cosmic variance due to finite simulation volume, and the third term denotes instrumental noise dominated by the system temperature \cite{Pober:2012zz}.
{The instrumental noise is related to the square of system temperature, i.e., $\sigma_{\rm ins} \propto T_{\rm sys}^{2}$, while the system temperature can be estimated by \cite{Pober:2012zz}
\begin{equation}
    T_{\rm sys}(\nu) = 16.3 \times 10^6 \, {\rm K} \left(\frac{\nu}{2\, \rm MHz}\right)^{-2.53} \ .
    \label{eq:foregroundfit}
\end{equation}
In this work, we adopt this system temperature in package \texttt{21cmSense} \cite{Pober:2012zz} to estimate the instrumental noise of \ac{SKA}.}
We adopt stations in the central area of the SKA1-low (an array of $295$ stations, where the diameter of each station is $35$\,m, distributed across a circular area with a diameter of $1.7$\,km), observing with a total bandwidth of $8\,\rm MHz$ and spectral resolution of $100\,\rm kHz$, and a total integration time of $10,000$ hours \cite{Braun:2019gdo}.
Marginalized uncertainty for a specific parameter $\theta_{i}$ satisfies $\sigma_{\theta_{i}} \geq \sqrt{(F^{-1})_{ii}}$ \cite{frechet1943extension}.
This suggests that the Fisher matrix provides conservative lower bounds on parameter constraints.
The $1\sigma$ uncertainty of the parameter $\theta_{i}$ is its standard deviation.
Furthermore, the correlation between parameters $\theta_{i} $ and $\theta_{j}$ is quantified by the dimensionless correlation coefficient $R_{ij} = C_{ij}/\sqrt{ C_{ii} C_{jj} }$, where the covariance matrix $C_{ij}$ is the inverse of the Fisher matrix $F_{ij}$.

\begin{figure*}
    \centering
    \includegraphics[width=0.8\textwidth]{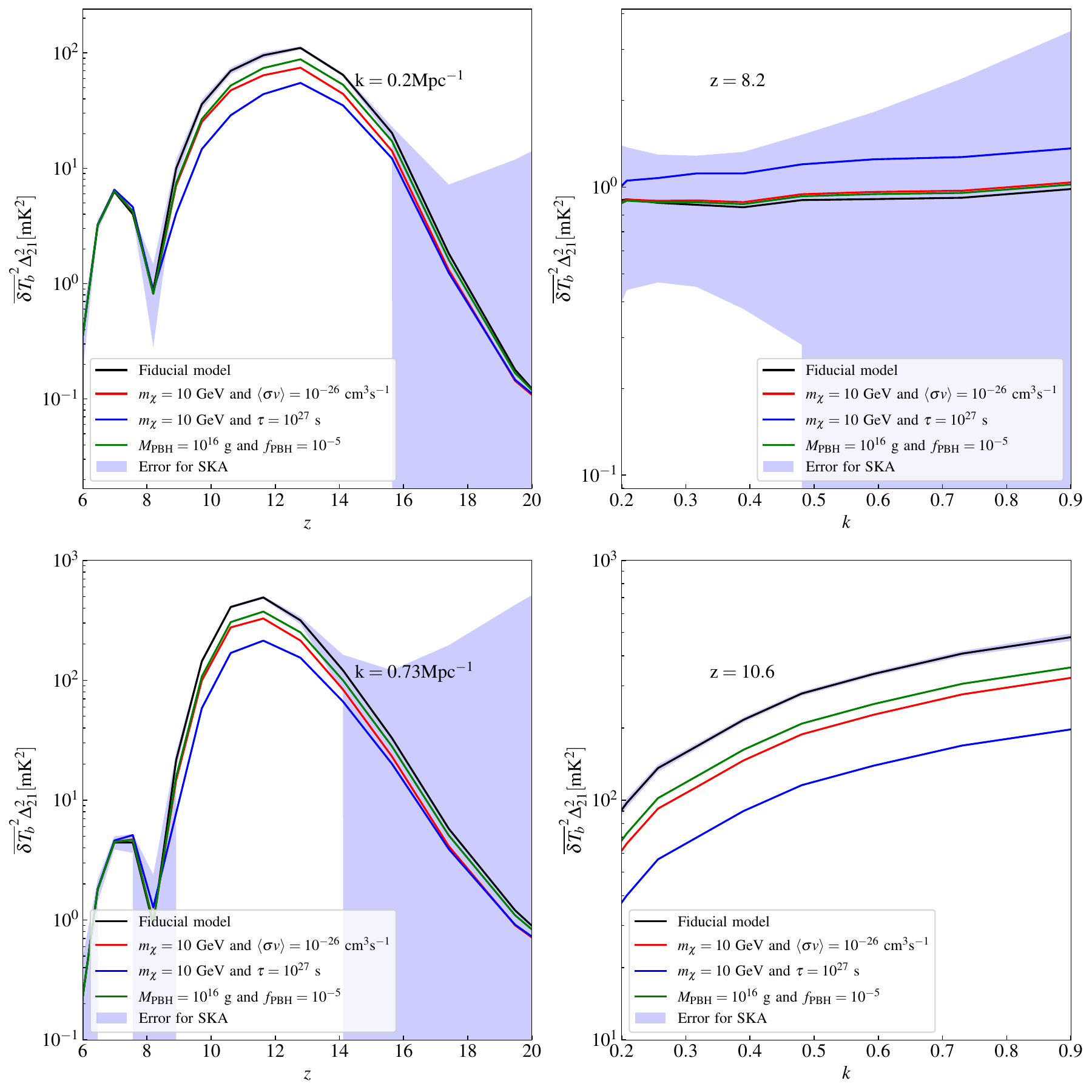}
    \caption{{The 21 cm power spectrum under different energy injection scenarios.
Left panels: The 21 cm power spectrum as a function of redshift $z$ at fixed scales $k = 0.2 \,\rm Mpc^{-1}$ and $k = 0.73\,\rm Mpc^{-1}$.
Right panels: The 21 cm power spectrum as a function of the scale at redshift $z = 8.2$ and $z = 10.6$.
In each panel, the instrumental noise is shown by the shaded region.
Black curves show the 21 cm power spectrum of the fiducial model.
Red curves show 21 cm power spectrum with \ac{DM} particle annihilation through $\chi\chi\rightarrow e^{+}e^{-}$ channel with $ m_{\chi} = 10 ~\rm{GeV}$ and $\langle \sigma v \rangle = 10^{-26}\, \rm{cm}^{3}\,s^{-1}$.
Blue curves show 21 cm power spectrum with \ac{DM} particle decay through $\chi\rightarrow e^{+}e^{-}$ channel with $ m_{\chi} = 10 ~\rm{GeV}$ and $\tau = 10^{27}\,\rm{s}$.
Green curves show 21 cm power spectrum with \ac{PBH} Hawking radiation, with $ M_{\rm PBH} = 10^{16} ~\rm{g}$ and $f_{\rm PBH} = 10^{-5}$.}}
    \label{fig:measurement error power}
\end{figure*}

Our model incorporates two categories of independent parameters, which are astrophysical parameters and parameters for \ac{DM} particles and \acp{PBH}.
Astrophysical parameters follow the conventions of \texttt{21cmFAST}, including $t_{\star}$, $a_{\star}$, $a_{\rm esc}$, ${\rm log}_{10}f_{\star}$, ${\rm log}_{10}f_{\rm esc}$, and ${\rm log}_{10}L_{\rm X}$, where $t_{\star}$ is 
the dimensionless star formation timescale, $a_{\star}$ represents the power-law index of stellar-to-halo mass ratio, $a_{\rm esc}$ represents the power-law index of UV photon escape fraction, ${\rm log}_{10}f_{\star}$ represents the stellar-to-halo mass ratio, ${\rm log}_{10}f_{\rm esc}$ represents the UV escape fraction, and ${\rm log}_{10}L_{\rm X}$ represents X-ray luminosity per star formation rate in unit of $ \rm erg \cdot yr \cdot sec^{-1} M^{-1}_{\odot}$, where $M_{\odot}$ stands for the solar mass.
The remaining parameters correspond to \ac{DM} physics.
$\langle \sigma v \rangle$ characterizes the annihilation cross section of \ac{DM} particles, in units of $ \rm cm^{3}~s^{-1}$.
$\Gamma = \tau^{-1}$ represents the decay rate of \ac{DM} particles, in units of $\rm s^{-1}$.
$f_{\rm PBH}$ signifies the abundance of \acp{PBH}.
For the fiducial model in Fig.~\ref{fig:measurement error power}, the astrophysical parameters are set to $t_{\star} = 0.5$, $a_{\star} = 0.5$, $a_{\rm esc} = -0.5$, ${\rm log}_{10}f_{\star} = -1.3$, ${\rm log}_{10}f_{\rm esc} = -1.0$, and ${\rm log}_{10}L_{\rm X} = 40.0$, while the \ac{DM} and \ac{PBH} parameters, $\langle \sigma v \rangle$, $\Gamma$, and $f_{\rm PBH}$ are set to zero.

We demonstrate how the 21 cm power spectrum responds to exotic energy injections and present the \ac{SKA}'s measurement errors on it, as shown in Fig.~\ref{fig:measurement error power}.
The left panels display the redshift evolution of the 21 cm power spectrum at different scales, revealing peaks during cosmic dawn ($z \sim 10 - 15$) and the epoch of reionization ($z \sim 6 - 8$).
The peak during cosmic dawn is dominated by the heating and ionizing effects of the \ac{IGM}, rendering the amplitude of this peak highly sensitive to energy injection processes such as \ac{DM} particle annihilation, decay, and \ac{PBH} Hawking radiation.
During cosmic dawn, exotic heating increases the \ac{IGM} kinetic temperature, thereby suppressing the 21 cm power spectrum amplitude.
Subsequently, after heating saturation, rising ionization causes the cosmic dawn peak to diminish.
Conversely, the peak during the epoch of reionization is governed by the ionizing effect of the \ac{IGM}.
According to Ref.~\cite{Mesinger:2013nua}, an increased ionization fraction amplifies the power spectrum during reionization.
This opposing response stems from distinct physical mechanisms, ionization reduces neutral hydrogen density during cosmic dawn, but enhances fluctuations in the ionized bubble during reionization.
The right panels show the scale dependence of the 21 cm power spectrum at fixed redshifts.
Exotic energy injections enhance the amplitude of the power spectrum at $z = 8.2$, but suppress it at $z = 10.6$, consistent with Ref.~\cite{Mesinger:2013nua}.
Crucially, near the redshift of the cosmic dawn peak, both the amplitude of the 21 cm power spectrum and the corresponding signal-to-noise ratio increase significantly.
Consequently, cosmic dawn emerges as the optimal observational window for the \ac{SKA} to probe \ac{DM}.

\section{Discovering potential of SKA}\label{sec:skadp}

In this section, we present the projected sensitivity of the \ac{SKA} for probing \ac{DM} particles and \acp{PBH}.
We further compare this sensitivity with existing constraints from astrophysical probes.

\subsection{Results for DM particles}

\begin{figure*}
    \centering
    \includegraphics[width=0.8\linewidth]{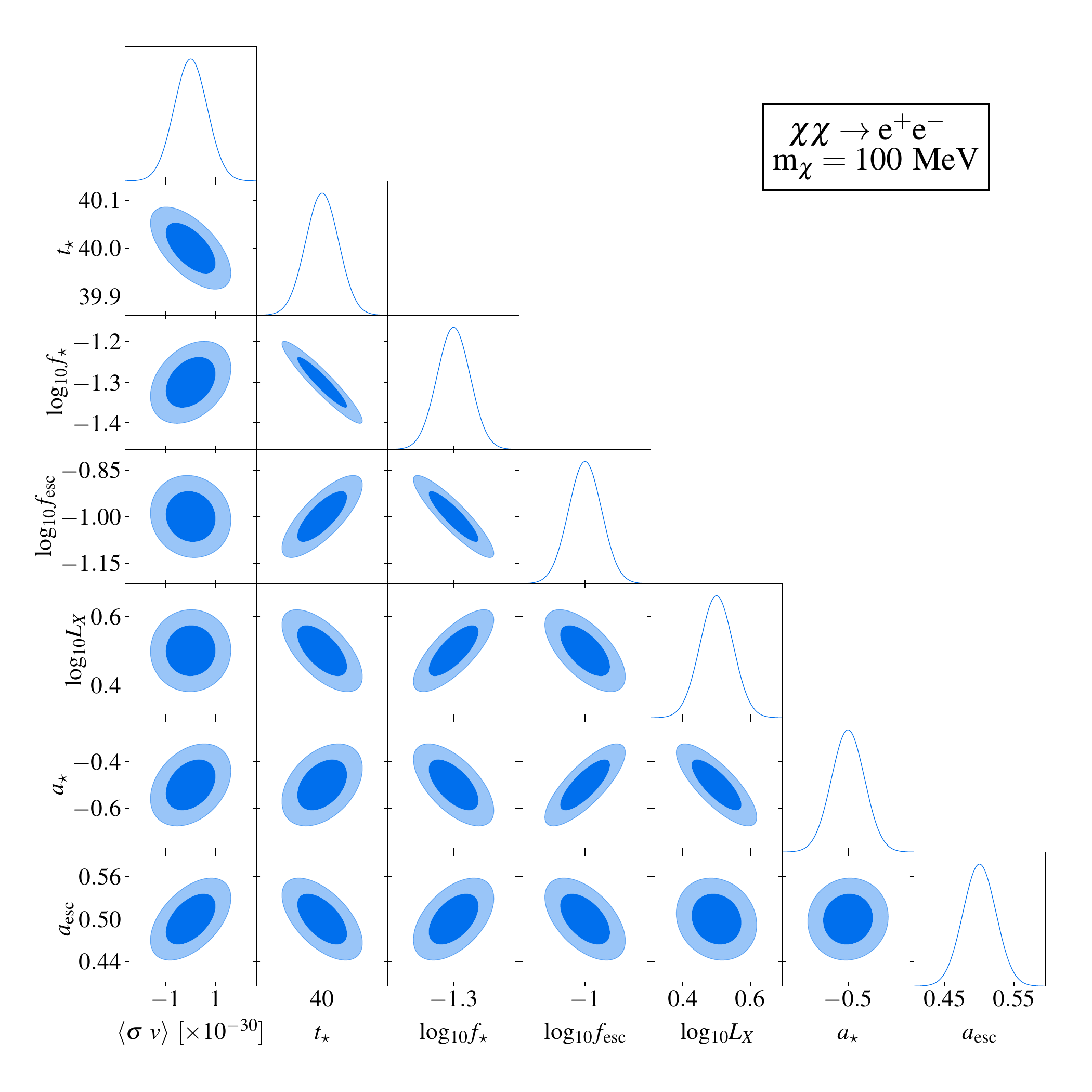}
    \caption{Fisher forecast for probing \ac{DM} annihilation through the $\chi \chi \rightarrow e^{+} e^{-}$ channel using the 21 cm power spectrum by the \ac{SKA}.
    $1\sigma$ and $2\sigma$ confidence intervals are represented by dark and light shaded areas, respectively, with solid curves indicating the marginalized posteriors.
    Fiducial model used is consistent with that shown in Fig.~\ref{fig:measurement error power}.
    The assumed \ac{DM} particle mass is $m_{\chi} = 100$\,MeV, integrated over $10,000$\,hours.}
    \label{fig:DM_ann_corr}
\end{figure*}

\begin{figure*}
    \centering
    \includegraphics[width=0.45\linewidth]{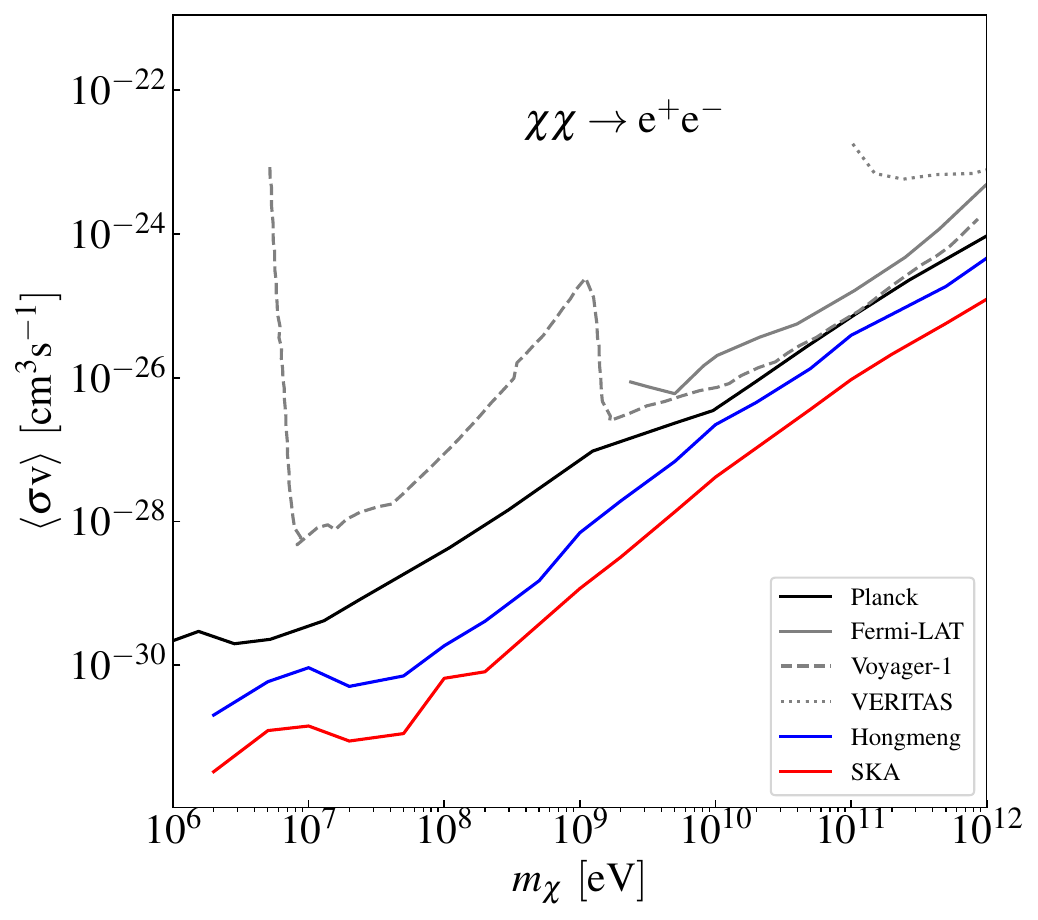}
    \includegraphics[width=0.45\linewidth]{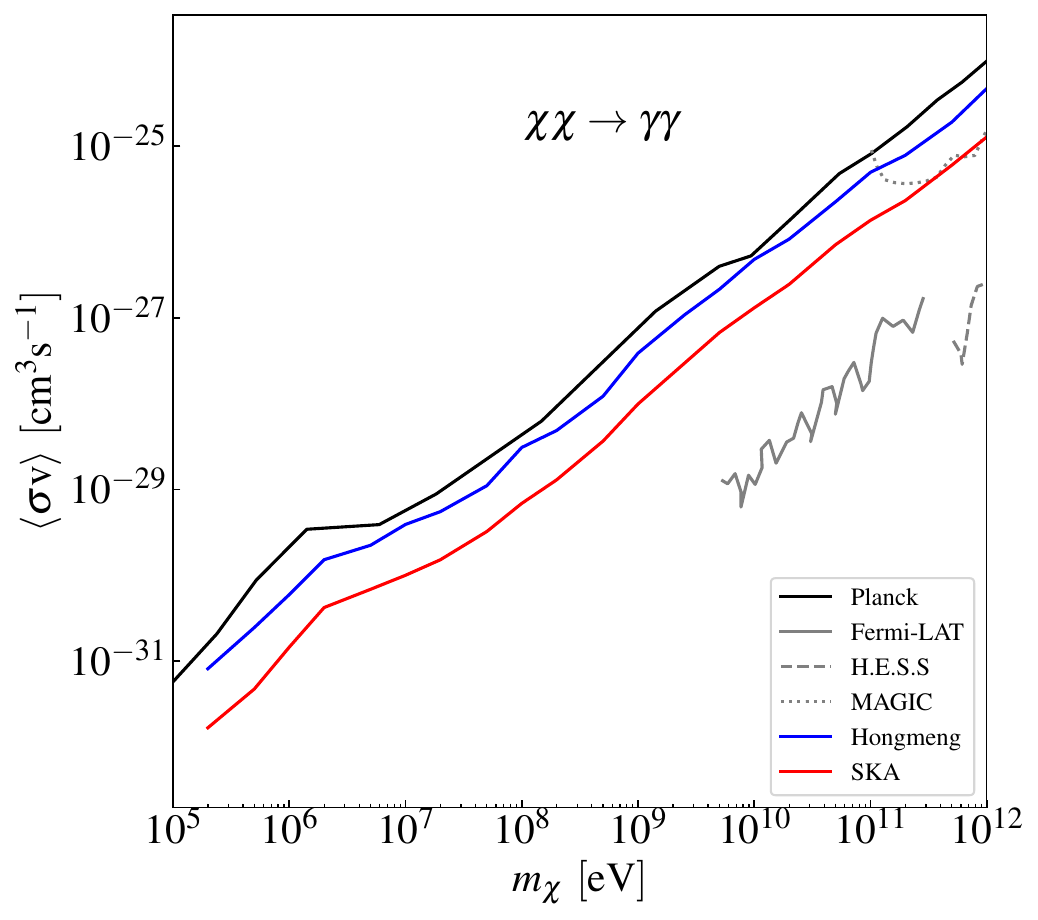}
    \includegraphics[width=0.45\linewidth]{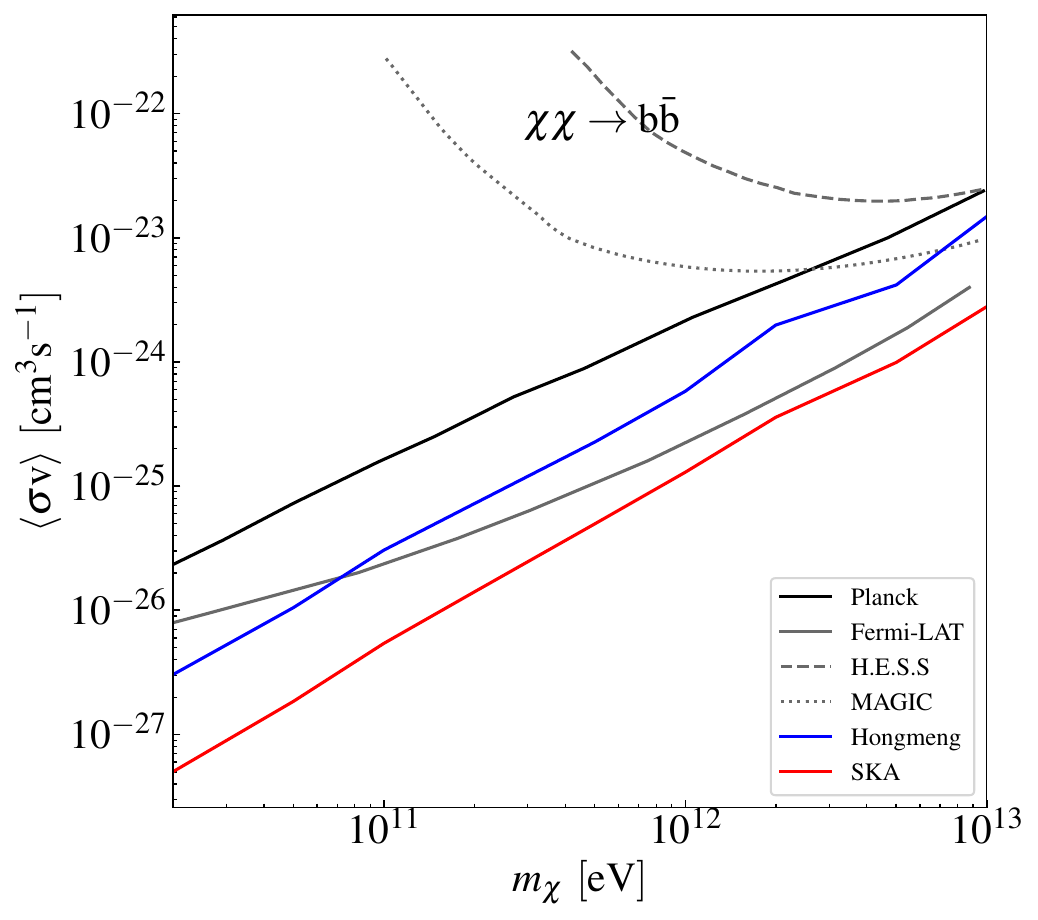}
    \caption{Prospective sensitivity of the \ac{SKA} for probing the annihilation of \ac{DM} particles through three channels.
    The $1\sigma$ confidence-level sensitivity of the \ac{SKA} to the thermally averaged annihilation cross section of \ac{DM} particles (mass range $10^{6}$-$10^{12}\,\text{eV}$) are shown by the red curves.
     Existing $2\sigma$ upper limits from observations of \ac{CMB} distortion (black curve) \cite{Zhang:2023usm}, gamma-ray observations (gray curves) \cite{Cirelli:2020bpc,HESS:2018kom,HESS:2014zqa,VERITAS:2017tif,MAGIC:2017avy,Aleksic:2013xea,Fermi-LAT:2015att}, and electron-positron pairs (gray dashed curve) \cite{Cohen:2016uyg,Boudaud:2018oya} are included for comparison.
    Prospective sensitivity of the 21 cm global spectrum (blue curve) \cite{Zhao:2024jad} is also included for comparison. }
    \label{fig:DM_ann_sense}
\end{figure*}

\begin{figure*}
    \centering
    \includegraphics[width=0.8\linewidth]{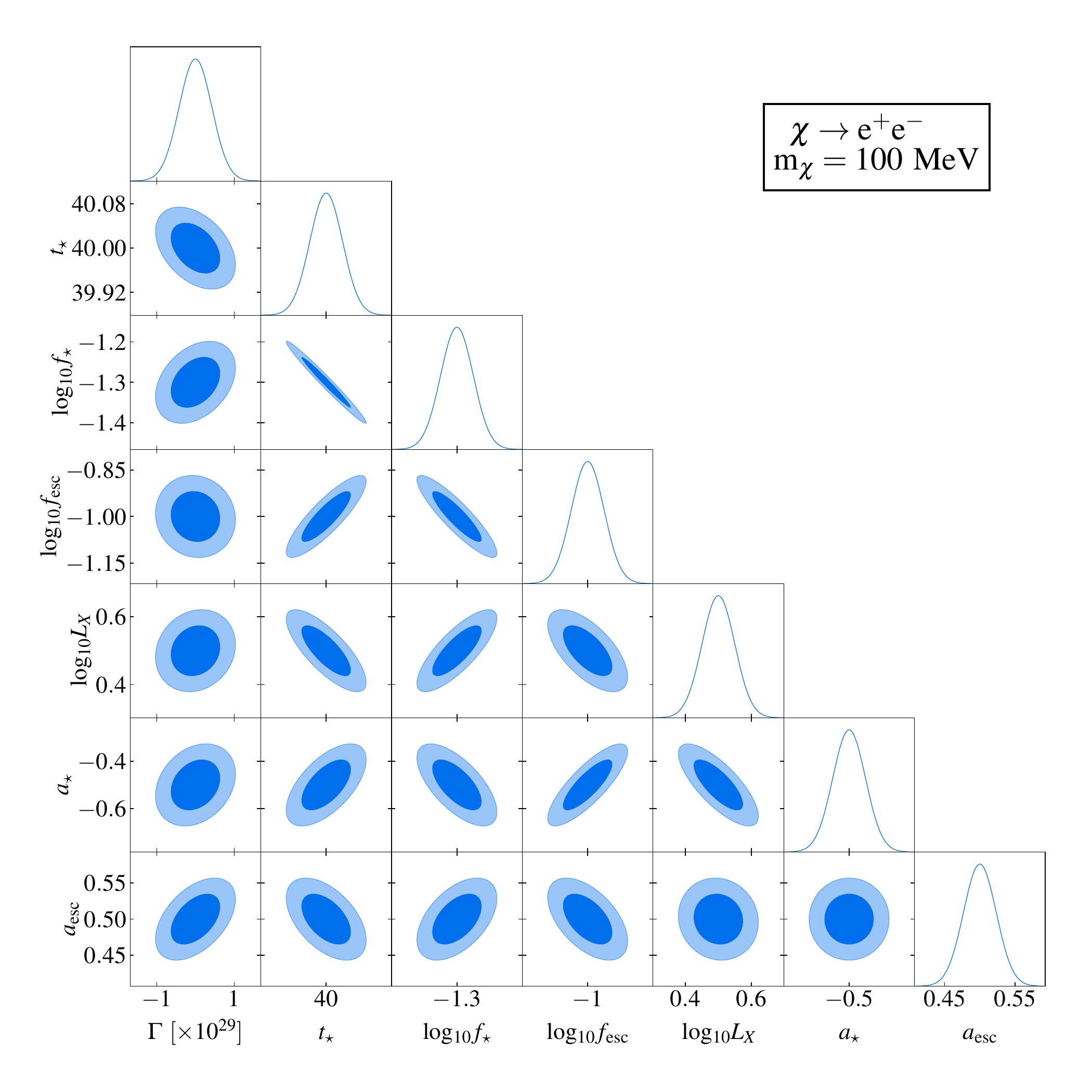}
    \caption{Fisher forecast for probing \ac{DM} decay through the $\chi \rightarrow e^{+} e^{-}$ channel using the 21 cm power spectrum by the \ac{SKA}.
    $1\sigma$ and $2\sigma$ confidence intervals are represented by dark and light shaded areas, respectively, with solid curves indicating the marginalized posteriors.
    Fiducial model used is consistent with that shown in Fig.~\ref{fig:measurement error power}.
    The assumed \ac{DM} particle mass is $m_{\chi} = 100$\,MeV, integrated over $10,000$\,hours.}
    \label{fig:DM_dec_corr}
\end{figure*}

\begin{figure*}
    \centering
    \includegraphics[width=0.45\linewidth]{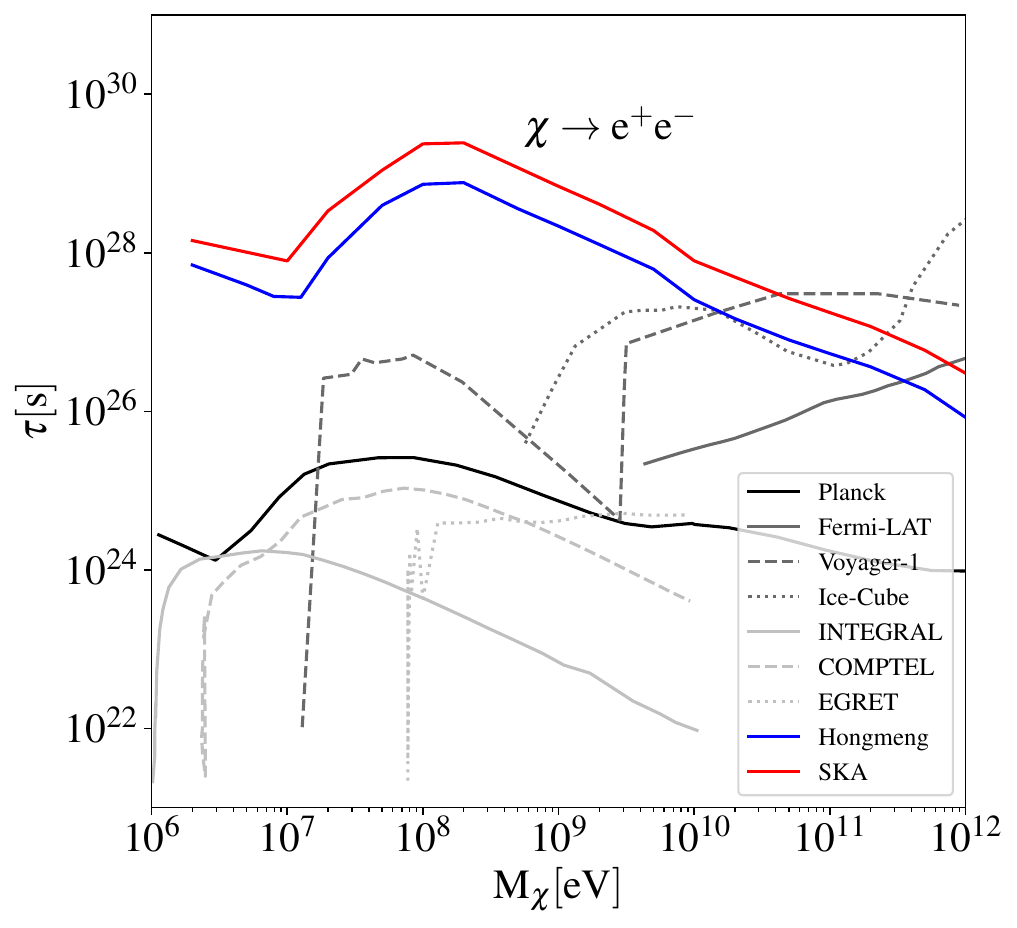}
    \includegraphics[width=0.45\linewidth]{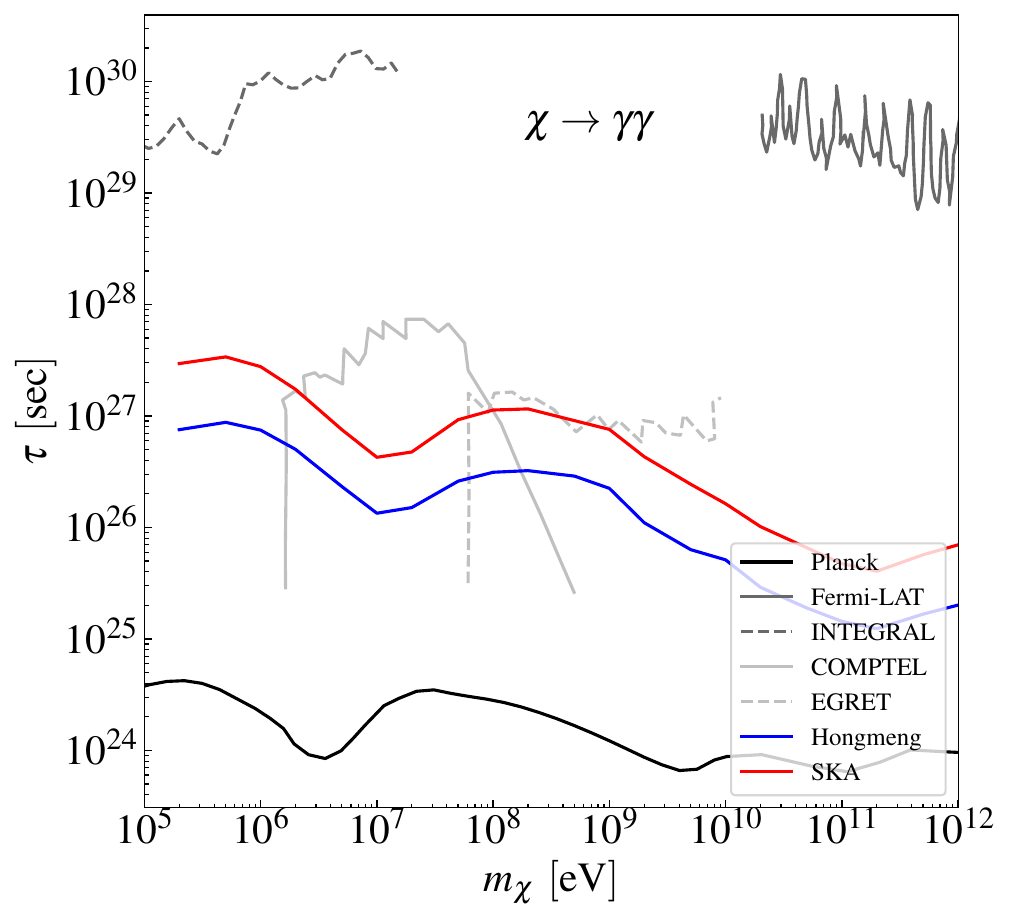}
    \includegraphics[width=0.45\linewidth]{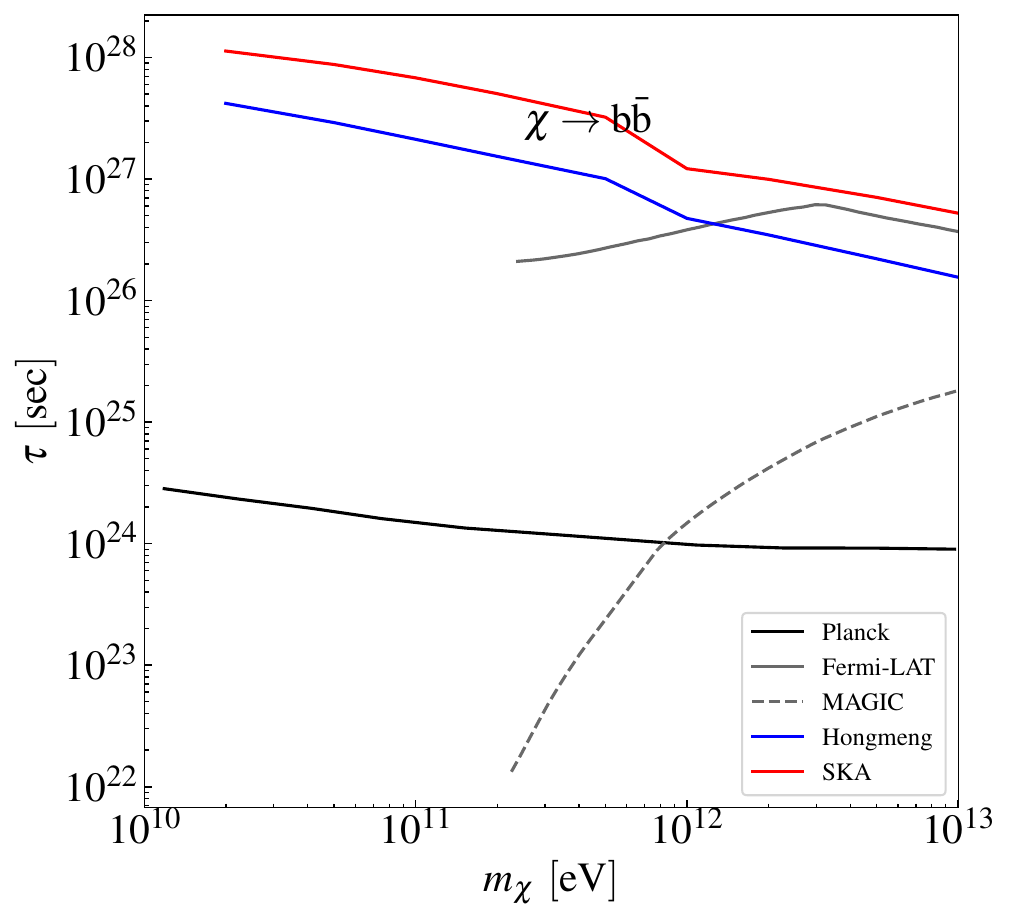}
    \caption{Prospective sensitivity of the \ac{SKA} for probing the decay of \ac{DM} particles through three channels.
    The $1\sigma$ confidence-level sensitivity of the \ac{SKA} project to the decay lifetime of \ac{DM} particles (mass range $10^{6}$-$10^{12}\,\text{eV}$) are shown by the red curves.
   Existing $2\sigma$ upper limits from observations of \ac{CMB} distortion (black curve) \cite{Planck:2018vyg,Capozzi:2023xie}, extragalactic photons (gray curves) \cite{Koechler:2023ual,Calore:2022pks,Foster:2022nva,Cirelli:2020bpc,Massari:2015xea,Cohen:2016uyg,Essig:2013goa}, and electron-positron pairs (gray dashed curve) \cite{Cohen:2016uyg,Boudaud:2018oya} are included for comparison.
    Prospective sensitivity of the 21 cm global spectrum (blue curve) is also taken into consideration \cite{Zhao:2024jad}. }
    \label{fig:DM_dec_sense}
\end{figure*}

The results of the Fisher information matrix analysis are summarized in Figs.~\ref{fig:DM_ann_corr}--\ref{fig:DM_dec_sense}.
Figs.~\ref{fig:DM_ann_corr} and \ref{fig:DM_dec_corr} display two
representative corner plots that characterize parameter correlations and constraints, assuming a fixed \ac{DM} particle mass of $100$\,MeV and an integration time of $10,000$\,hours.
The $1\sigma$ and $2\sigma$ confidence intervals are represented by the dark and light shaded areas, respectively, while the solid curves depict the marginalized posterior distributions.
Fiducial model parameters in these figures are consistent with those in Fig.~\ref{fig:measurement error power}.
Complete corner plots are provided in Appendix \ref{sec:appendix}.
Figs.~\ref{fig:DM_ann_sense} and \ref{fig:DM_dec_sense} quantify the projected sensitivity at $1\sigma$ confidence level for \ac{SKA}'s capability to probe \ac{DM} and compare these results with existing constraints at $2\sigma$ confidence level from \ac{CMB} observations \cite{Zhang:2023usm}, gamma ray measurements \cite{Cirelli:2020bpc,HESS:2018kom,VERITAS:2017tif,MAGIC:2017avy,Fermi-LAT:2015att,HESS:2014zqa,Aleksic:2013xea}, electron-positron pair observations \cite{Boudaud:2018oya,Boudaud:2016mos,Cohen:2016uyg}, and 21 cm global spectrum measurements \cite{Zhao:2024jad}.

Figs.~\ref{fig:DM_ann_corr} and \ref{fig:DM_dec_corr} reveal mild correlations between the annihilation and decay parameters of \ac{DM} particles and the astrophysical parameters.
This result indicates a limited degeneracy between the \ac{DM}-induced exotic energy injection and astrophysical effects on the 21 cm power spectrum.
Specifically, $\langle \sigma v \rangle$ and $\tau$ show weak positive correlations with $a_{\rm esc}$, $a_{\star}$ and $\log_{10} f_{\star}$, mild negative correlations with $t_\star$, and negligible correlations with $L_{\rm X}$ and $f_{\rm esc}$.
These results demonstrate the feasibility of constraining \ac{DM} parameters and extracting key properties, particularly the thermally-averaged annihilation cross section $\langle \sigma v \rangle$ and decay lifetime $\tau$, using the 21 cm power spectrum.
The weak degeneracies demonstrate that the 21 cm power spectrum can effectively constrain \ac{DM} parameters independently of astrophysical uncertainties.
This allows probing of fundamental \ac{DM} properties, particularly $\langle \sigma v \rangle$ and $\tau$, with minimized contamination from astrophysical processes.

Fig.~\ref{fig:DM_ann_sense} demonstrates \ac{SKA}'s sensitivity to constrain \ac{DM} annihilation via the $\chi\chi \rightarrow e^{+}e^{-}$ channel is superior to that for other channels such as $\chi\chi \rightarrow \gamma\gamma$ and $\chi\chi \rightarrow b\bar{b}$.
Focusing on the optimal annihilation channel $\chi\chi \rightarrow e^{+}e^{-}$ (upper panel), we find that utilizing the \ac{SKA}, with $10,000$ hours of integration time, the 21 cm power spectrum can achieve a sensitivity of $\langle\sigma v\rangle \leq 10^{-28}\,{\rm cm}^{3}\,{\rm s}^{-1}$ for $10$\,GeV dark matter particles.
This sensitivity surpasses the most stringent current constraints (gray curves), demonstrating \ac{SKA}'s capacity to test existing limits in the near future.
Furthermore, the \ac{SKA} exhibits superior sensitivity to sub-GeV dark matter, a mass range where conventional probing experiments provide only weak constraints.
While extended integration would improve sensitivity, practical implementation faces instrumental stability challenges \cite{Braun:2019gdo}.

The blue curve in Fig.~\ref{fig:DM_ann_sense} shows our previous result based on the Hongmeng project, which assessed the capability of constraining \ac{DM} annihilation parameters using the 21 cm global spectrum.
In this work, we compare the results obtained by the \ac{SKA} with those from Hongmeng.
Our analysis reveals that the 21 cm power spectrum exhibits weaker correlations between \ac{DM} parameters and astrophysical parameters than the 21 cm global spectrum.
This reduced degeneracy enables the power spectrum to extract \ac{DM}-induced signals more effectively. 
With the same integration time, the \ac{SKA} achieves higher sensitivity than that implied by the results obtained by Hongmeng with the global spectrum, demonstrating its potential to probe \ac{DM} annihilation signals beyond the reach of the global spectrum in the near future.
Additionally, unlike the 21 cm global spectrum, the 21 cm power spectrum contains information on different scales.
Therefore, by probing the 21 cm power spectrum, the \ac{SKA} is expected to provide insights into the properties of \ac{DM} by measuring its effects on different scales, thus deepening our understanding of the \ac{DM} nature.

{Furthermore, a comparison between the power spectrum and the global spectrum from the \ac{SKA} could provide more comprehensive insights.
However, a detailed discussion of the Fisher matrix analysis and noise modeling for the global spectrum is beyond the scope of this paper.
Therefore, we only present a qualitative analysis here, and defer a more thorough investigation to our upcoming work.
Based on our previous work, constraints on \ac{DM} derived from the global spectrum depend on its measurement error, which comprises two noise components: the foreground residual and the instrumental noise.
In this work, the blue curve corresponds to a scenario with 10,000 hours of observation time and a foreground residual level of 0.001.
This configuration has already been examined in our previous work, in which the instrumental noise and foreground residual are comparable \cite{Zhao:2024jad}.
Changing the telescope only affects the instrumental noise, while the foreground noise remains unchanged.
Assuming the same observational setup, i.e., 10,000 hours and 0.001 foreground residual, we can qualitatively analyze the measurement error of the \ac{SKA}.
Qualitatively, for a fixed integration time, the instrumental noise is approximately inversely proportional to the effective collecting area of the telescope.
Given that the \ac{SKA} has a much larger effective area than the spectrometer employed by the Hongmeng project, its instrumental noise is expected to be significantly lower for the same observation time.
Thus, the instrumental noise of the \ac{SKA} would be lower than the foreground residual, and the total noise would be dominated by the latter.
Consequently, under the assumptions of 10,000 hours of integration and 0.001 foreground residual, observations with the \ac{SKA} are not expected to yield better constraints than those from Hongmeng, which are represented by the blue curves.
A more detailed quantitative analysis is beyond the aim of this paper. We plan to carry out a thorough and precise analysis in a forthcoming work.
}

Fig.~\ref{fig:DM_dec_sense} indicates that the \ac{SKA} has superior sensitivity in constraining \ac{DM} decay via the $\chi \rightarrow e^{+}e^{-}$ channel, compared to alternative channels such  $\chi \rightarrow \gamma\gamma$ and $\chi \rightarrow b\bar{b}$.
Focusing on the optimal decay channel $\chi \rightarrow e^{+}e^{-}$ (upper panel), and assuming an integration time of $10,000$\,hours, the \ac{SKA} is projected to improve constraints on the \ac{DM} particle decay lifetime by two orders of magnitude, surpassing current experimental bounds (gray curves).
This result demonstrates \ac{SKA}'s potential to test existing \ac{DM} decay models in the near future.
While increasing the integration time would further improve sensitivity, practical implementation beyond $10,000$\,hours may require addressing instrumental stability limitations \cite{Braun:2019gdo}.
Notably, through 21 cm power spectrum measurements, the \ac{SKA} achieves superior sensitivity to sub-GeV \ac{DM}, a parameter space weakly constrained by current methods.

The blue curve in Fig.~\ref{fig:DM_dec_sense} represents our previous work, which explored the potential to probe \ac{DM} decay using the 21 cm global spectrum from the Hongmeng Project.
We perform a comparative analysis of \ac{DM} decay constraints derived from the 21 cm power spectrum and the 21 cm global spectrum.
Our analysis reveals less degeneracy between \ac{DM} decay parameters and astrophysical parameters in the power spectrum compared with the global spectrum.
This reduced degeneracy enables the power spectrum to more effectively extract \ac{DM}-induced signatures.
With the same integration time, the 21 cm power spectrum achieves a sensitivity one order of magnitude better than 21 cm global spectrum.
This result demonstrates that the \ac{SKA} will impose tighter constraints on \ac{DM} decay parameter than those achieved with the 21 cm global spectrum.
Furthermore, unlike the global spectrum, the 21 cm power spectrum encodes information across multiple spatial scales, enabling the probing of \ac{DM} properties at different scales and deepening our understanding of \ac{DM}’s fundamental nature.

\subsection{Results for PBHs}

\begin{figure*}
    \centering
    \includegraphics[width=0.8\linewidth]{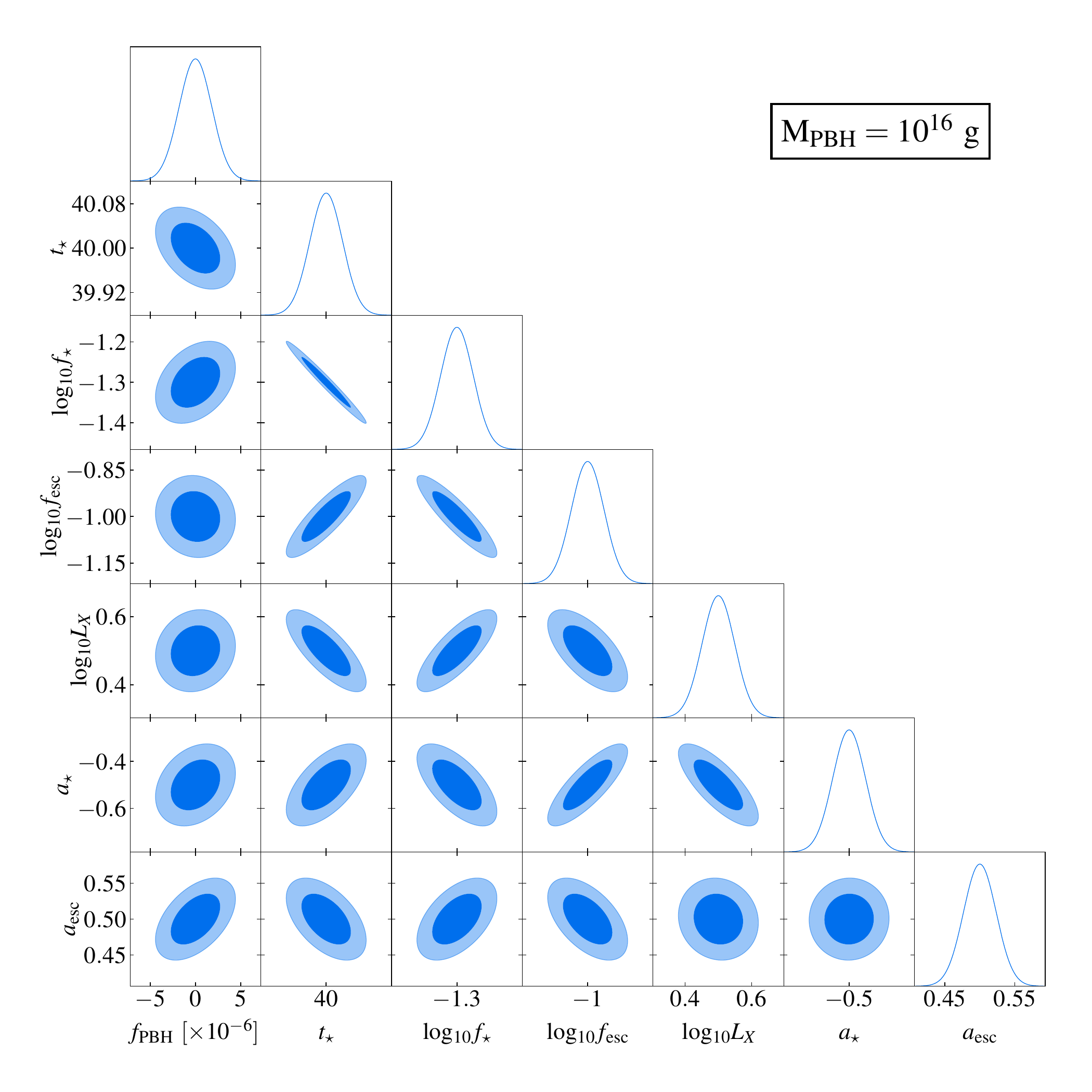}
    \caption{Fisher forecast for probing \ac{PBH} Hawking radiation using the 21 cm power spectrum by the \ac{SKA}.
    Dark and light shaded regions correspond to contours at $1\sigma$ and $2\sigma$ confidence intervals, respectively.
    Solid curves represent the marginalized posteriors of the model parameters.
    Fiducial model used is consistent with that shown in Fig.~\ref{fig:measurement error power}.
    The mass of \ac{PBH} is assumed to be $M_{\rm PBH} = 10^{16}$\,g, with an integration duration of $10,000$ hours.}
    \label{fig:PBH_corr}
\end{figure*}

\begin{figure*}
    \centering
    \includegraphics[width=0.6\linewidth]{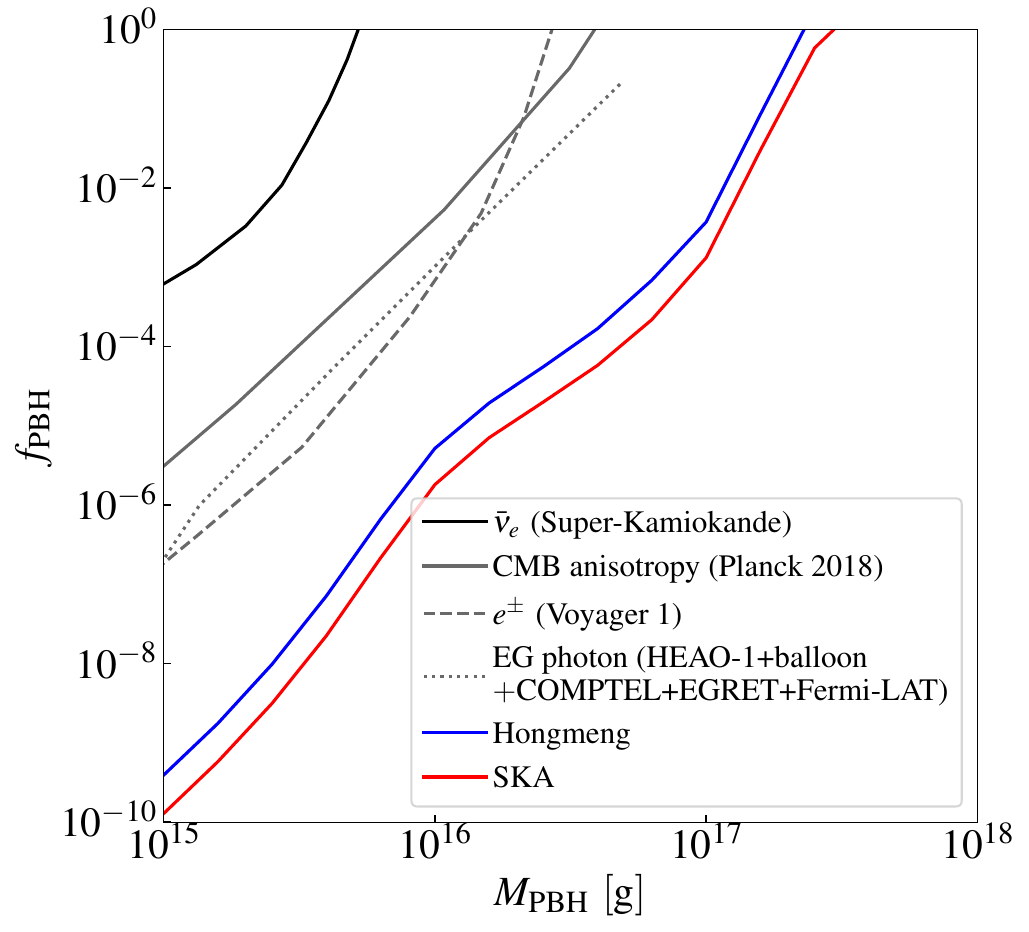}
    \caption{Prospective sensitivity of the \ac{SKA} for probing the \acp{PBH}.
    The $1\sigma$ confidence-level sensitivity of the \ac{SKA} project to measure the abundance of \acp{PBH} within the mass range of $10^{15}-10^{18}$\,g is shown by the red curve.
    For comparison, we show the existing upper limits at $2\sigma$ confidence level from observations of the diffusion neutrino background (black curve) \cite{Wang:2020uvi}, \ac{CMB} anisotropies (gray solid curve) \cite{Chluba:2020oip,Acharya:2020jbv,Clark:2016nst}, extra-galactic photons (gray dotted curve) \cite{Carr:2016hva}, and electron-positron pairs (gray dashed curve) \cite{Boudaud:2018hqb}. 
    Prospective sensitivity of the 21 cm global signal (blue curve) is also taken into consideration.}
    \label{fig:PBH_sense}
\end{figure*}

Figs.~\ref{fig:PBH_corr} and \ref{fig:PBH_sense} summarize the results of our parameter estimation based on Fisher matrix analysis.
Fig.~\ref{fig:PBH_corr} shows the correlations and constraints of model parameters for a $10^{16}$\,g \ac{PBH} with an integration time of $10,000$\,hours.
Shaded regions represent the $1\sigma$ (dark) and $2\sigma$ (light) confidence regions.
Corresponding one-dimensional marginalized posterior distributions are shown as solid curves.
Fiducial model parameters are consistent with those in Fig.~\ref{fig:measurement error power}.
Fig.~\ref{fig:PBH_sense} quantifies the \ac{SKA}’s sensitivity to Hawking radiation for \acp{PBH} with mass ranging from $10^{15}$\,g to $10^{18}$\,g, assuming a fixed integration time of $10,000$\,hours, at the $1\sigma$ confidence level.
This sensitivity is compared with existing $2\sigma$ exclusion bounds from observations of the diffuse neutrino background \cite{Wang:2020uvi}, \ac{CMB} anisotropy \cite{Chluba:2020oip,Acharya:2020jbv,Clark:2016nst}, gamma ray measurements \cite{Carr:2016hva}, electron-positron pair measurements \cite{Boudaud:2018hqb,Cohen:2016uyg}, and 21 cm global signals \cite{Zhao:2024jad}.

Fig.~\ref{fig:PBH_corr} reveals weak correlations between the \ac{PBH} abundance $f_{\rm PBH}$ and key astrophysical parameters.
This result suggests weak degeneracies between the exotic energy injection from the \ac{PBH} Hawking radiation and astrophysical effects on the 21 cm power spectrum. Specifically, $f_{\rm PBH}$ exhibits weak positive correlations with $a_{\rm esc}$, $a_{\star}$, and $\log_{10}f_{\star}$, and a negative correlation with $t_{\star}$. In contrast, $f_{\rm PBH}$ shows no significant correlations with $L_{\rm X}$ and $\log_{10}f_{\rm esc}$.
These weak degeneracies enable the 21 cm power spectrum to constrain \ac{PBH} properties while minimizing contamination from astrophysical uncertainties.
This result allows the probing of fundamental \ac{PBH} properties, particularly the abundance $f_{\rm PBH}$.

Based on the results in Fig.~\ref{fig:PBH_sense}, we demonstrate that the 21 cm power spectrum measured by the \ac{SKA} achieves a sensitivity to $f_{\rm PBH} \simeq 10^{-10}$ for \acp{PBH} with masses of $10^{15}$\,g, assuming an integration time of $10,000$\,hours.
This result surpasses constraints from existing observations (gray curves) by up to $3-4$ orders of magnitude, indicating that the \ac{SKA} can test these results in the near future. 
Moreover, the \ac{SKA} extends sensitivity to higher-mass \acp{PBH} compared with current experiments, particularly probing the unexplored mass range above $10^{17}$\,g. 
While extended integration would improve sensitivity, practical operation beyond $10,000$\,hours may be limited by instrumental stability constraints \cite{Braun:2019gdo}.

Our prior analysis using Hongmeng's 21 cm global spectrum also established \ac{PBH} abundance constraints.
Such results are presented by the blue curve in Fig.~\ref{fig:PBH_sense}.
In this work, we compare the constraints on \acp{PBH} derived from the 21 cm power spectrum with those obtained from the 21 cm global spectrum.
Our result indicates a weaker coupling of \ac{PBH} abundance and astrophysical parameters in the 21 cm power spectrum than that in the global spectrum.
With the same integration time, the 21 cm power spectrum using the \ac{SKA} reaches a sensitivity nearly an order of magnitude higher than that achieved by the 21 cm global spectrum with Hongmeng.
This result suggests that, in the near future, the \ac{SKA} is expected to place tighter constraints on \ac{PBH} abundance than those obtained from the 21 cm global spectrum, thus allowing us to probe \acp{PBH} with unprecedented precision.

\section{Summary}\label{sec:sumdc}

In this work, we systematically assess the prospects for the \ac{SKA} to probe \ac{DM} particle annihilation, decay, and \ac{PBH} Hawking radiation via the 21 cm power spectrum during cosmic dawn.
Exotic energy from these \ac{DM} processes can deposit into \ac{IGM}, thereby heating and ionizing the \ac{IGM}, significantly suppressing the 21 cm power spectrum.
Utilizing Fisher matrix analysis, we quantify \ac{SKA}'s sensitivity to the relevant parameters, specifically the annihilation cross section ($\langle\sigma v\rangle$), decay lifetime ($\tau$), and \ac{PBH} abundance ($f_{\mathrm{PBH}}$).
Our results demonstrate that, with its designed observational capabilities, the \ac{SKA} is uniquely positioned to place breakthrough constraints on these parameters.

The \ac{SKA} demonstrates great observational capability in probing DM particle annihilation and decay.
Specifically, for \ac{DM} annihilation and decay via channels producing electron-positron pairs,  the \ac{SKA} achieves optimal sensitivity for extracting \ac{DM} signals using the 21 cm power spectrum during cosmic dawn.
With $10,000$ hours of integration time, the \ac{SKA} is projected to reach sensitivities of $\langle\sigma v\rangle\leq 10^{-28}~\text{cm}^{3}~\text{s}^{-1}$ and $\tau\geq 10^{28}~\text{s}$ for $10$\,GeV \ac{DM} particles, surpassing existing astrophysical constraints by approximately $2-3$ orders of magnitude.
Moreover, it provides tighter constraints than current experimental limits on sub-GeV \ac{DM} particles, deepening our understanding of \ac{DM} properties.

For \ac{PBH} Hawking radiation, the \ac{SKA} can constrain the \ac{PBH} abundance to $f_{\rm PBH}\leq 10^{-6}$ for \acp{PBH} with masses of $10^{16}$\,g and an integration time of $10,000$\,hours, representing a three-order-of-magnitude improvement over existing limits.
Crucially, the \ac{SKA} can probe more massive \acp{PBH} ($\geq 10^{17}$\,g), which are currently undetectable by other probes, thereby opening a new window to test \acp{PBH} as \ac{DM} candidates.

In summary, our analysis demonstrates that, during the cosmic dawn, the \ac{SKA} can effectively probe \ac{DM} annihilation, decay, and \ac{PBH} Hawking radiation through 21 cm power spectrum measurements with $10,000$ hours of integration time. 
The sensitivity of the \ac{SKA} is fundamentally limited by instrumental noise, which decreases with increasing integration time. 
However, while insufficient integration time degrades detection sensitivity, significantly extending integration time introduces practical challenges in maintaining instrumental stability, a systematic consideration beyond the scope of this study.

{It is noteworthy that in this work, we have not considered the scenario in which \ac{DM} particle annihilation, decay, and Hawking radiation from \acp{PBH} coexist. However, this represents an important scenario, which may provide valuable insights into the nature of \ac{DM}. Methods such as the Fisher information matrix may help break the degeneracies among these mixed \ac{DM} processes, but the relevant discussions are more complex and beyond the scope of this work. We anticipate conducting a more detailed investigation in our future work.}

{Furthermore, alternative methods such as the bispectrum, scattering transforms, and Minkowski functionals are also expected to provide additional constraining capability.
For example, the bispectrum captures higher-order statistic information, i.e., the non-Gaussianity that is generally absent in the analysis of power spectrum \cite{Chhabra:2025dfo,Gill:2025ydz,Noble:2024uzl}.
However, this method is computationally more complex than the power spectrum due to the necessity to compute the three-point correlation function.
Similarly, scattering transforms provide rich, high-order descriptions of signal morphology but require substantial computational resources to compute all scattering paths \cite{Diao:2025adp,Shimabukuro:2025equ,Zhao:2023tep}.
In contrast, Minkowski functionals are computationally efficient compared to the power spectrum, yet they offer limited scale-dependent information and are less sensitive to anisotropy \cite{Diao:2024wrf}.
Therefore, the power spectrum remains a relatively more cost-efficient approach compared with these methods.
Nevertheless, since each method has its own advantages and limitations, a combined approach may yield particularly valuable insights. We are thus preparing a new work that incorporates such methods.}

Looking beyond cosmic dawn, investigating the nature of \ac{DM} during the dark ages ($z \gtrsim 30$) represents a frontier in cosmology.
While recently the \ac{JWST} has detected high-redshift galaxies \cite{2025arXiv250612129L}, the dark ages remain observationally unexplored.
As galaxies have not yet formed during this epoch, 21 cm signal provides the only radio probe for the dark ages.
Crucially, unlike the 21 cm signal in cosmic dawn, the 21 cm signal in the dark ages is pristine, which is unaffected by astrophysical heating and ionization.
Therefore, targeting the dark ages with next-generation radio telescopes like the \ac{SKA} and the Hongmeng offers an effective pathway to probe fundamental physics of \ac{DM}.
Consequently, observations of dark ages circumvent the degeneracies between \ac{DM} physics and astrophysics at lower redshift, enabling stringent constraints on the parameters of \ac{DM}.
 In summary, 21 cm cosmology holds exceptional promise as a uniquely powerful probe for unveiling the physics of the early universe and the fundamental nature of \ac{DM}.

\appendix
\section{Results of Fisher matrix analysis}

\begin{figure*}
    \includegraphics[width=0.8\textwidth]{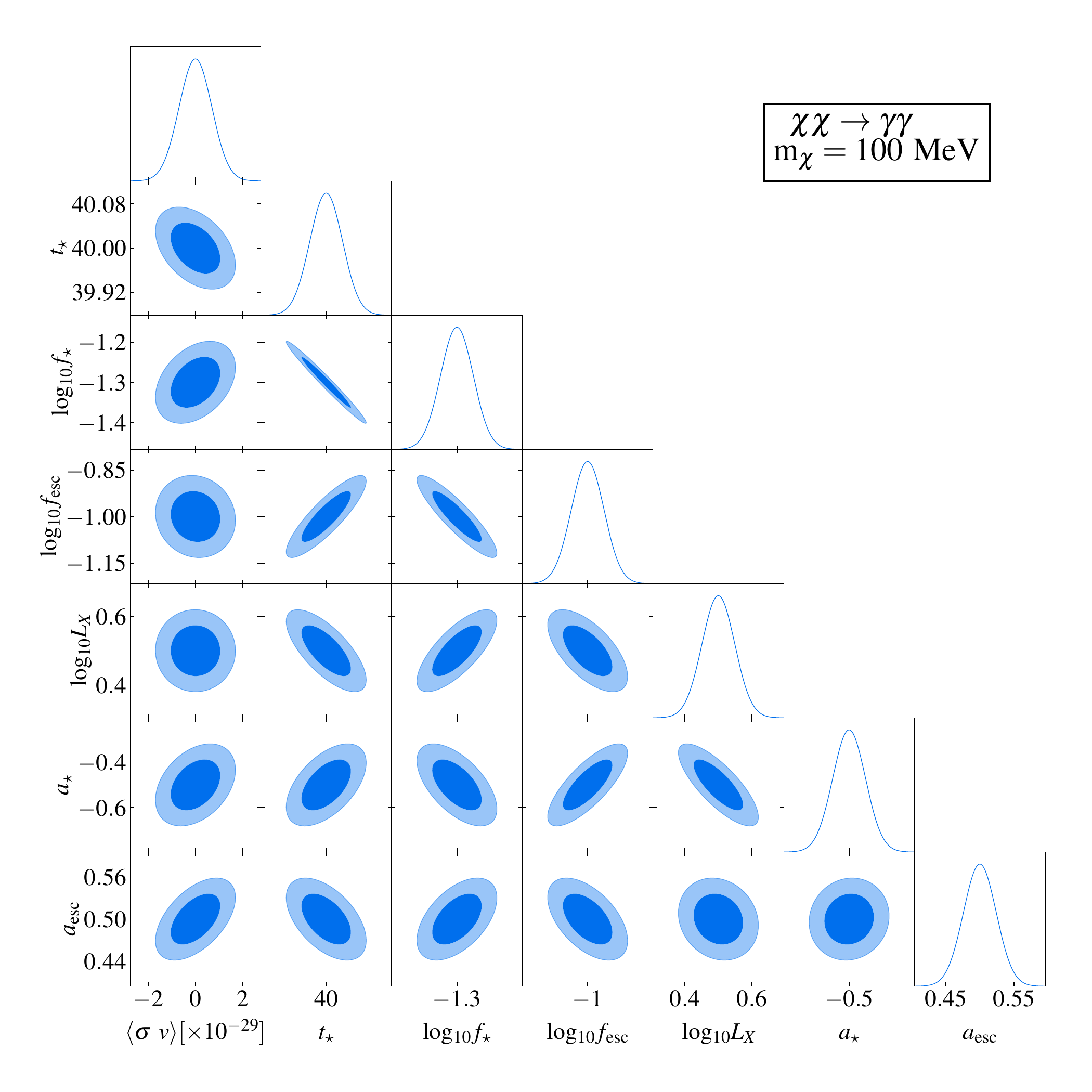}
    \caption{Fisher forecast for probing \ac{DM} annihilation through the $\chi \chi \rightarrow \gamma\gamma$ channel using 21 cm power spectrum by the \ac{SKA}.
    $1\sigma$ and $2\sigma$ confidence intervals are represented by dark and light shaded areas, respectively, with solid curves indicating the marginalized posteriors.
    Fiducial model used is consistent with that shown in Fig.~\ref{fig:measurement error power}.
    The assumed \ac{DM} particle mass is $m_{\chi} = 100$\,MeV, integrated over $10,000$\,hours.}
\label{fig:ann_corry}
\end{figure*}

\begin{figure*}
    \includegraphics[width=0.8\textwidth]{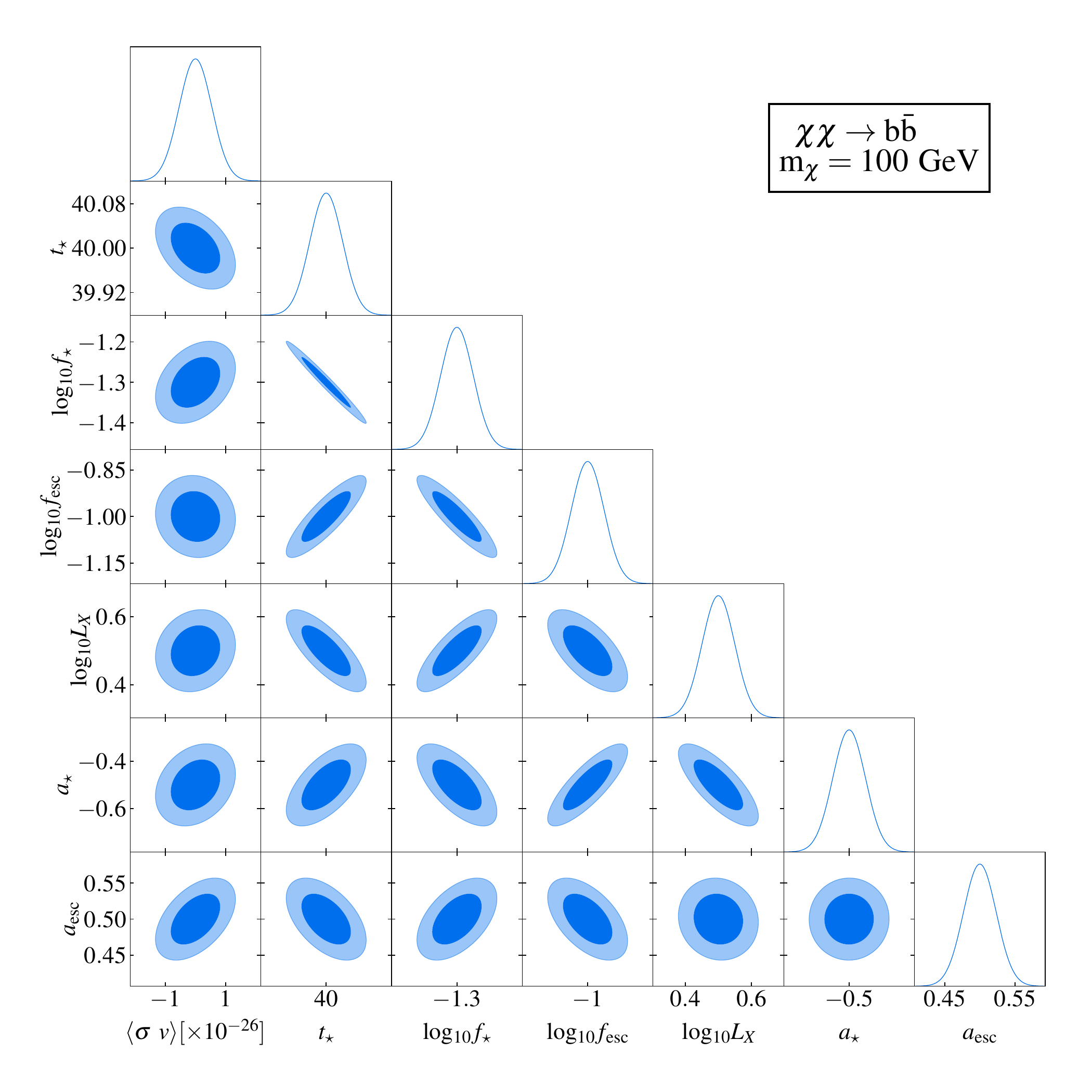}
    \caption{Same as Fig.~\ref{fig:ann_corry} but for $100$\,GeV \ac{DM} particles annihilating into bottom-anti-bottom quark pairs.
    }
    \label{fig:ann_corrb}
\end{figure*}

\begin{figure*}

\includegraphics[width=0.8\textwidth]{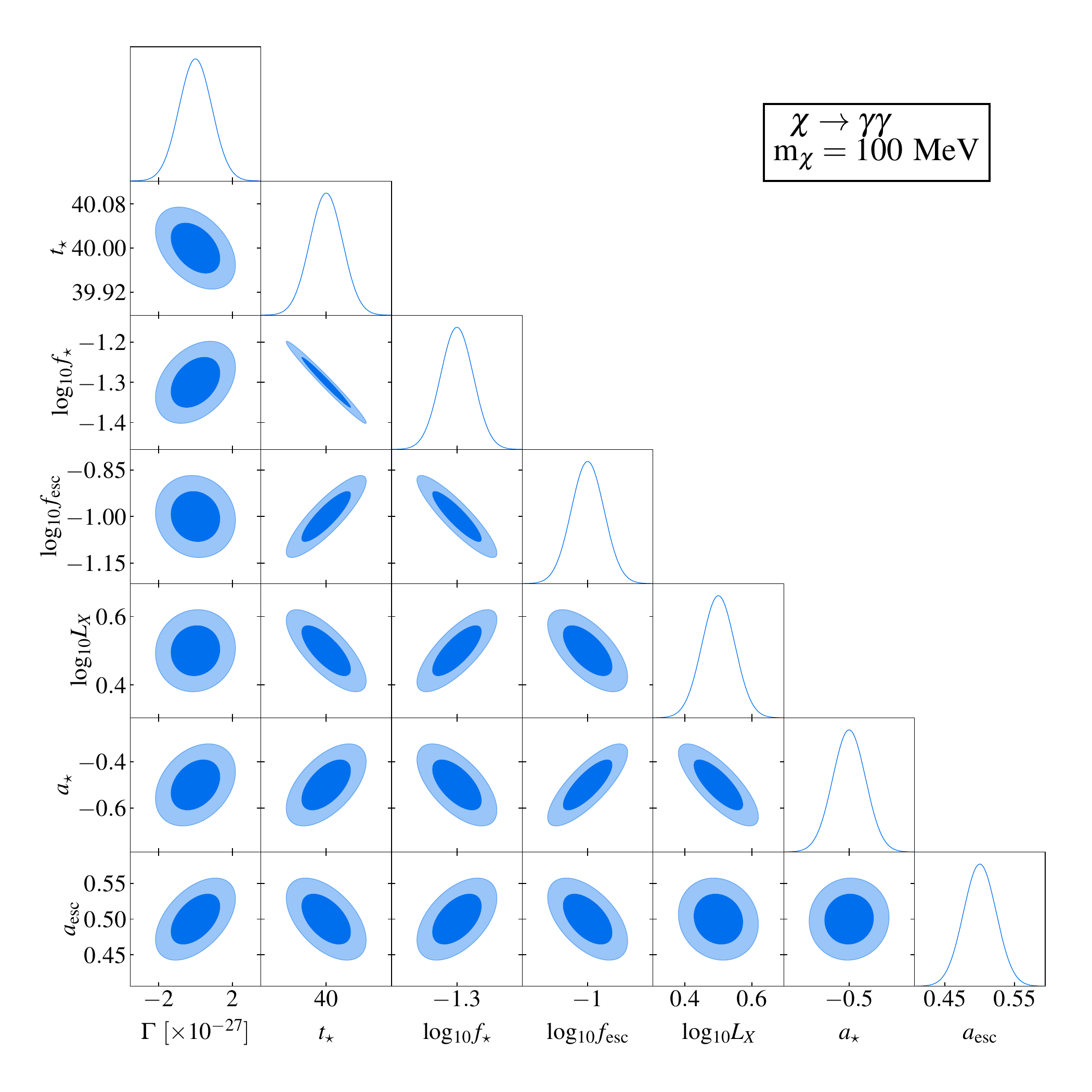}
\caption{Same as Fig.~\ref{fig:ann_corry} but for $100$\,MeV \ac{DM} particles decaying into photon pairs.
}
\label{fig:dec_corry}
\end{figure*}

\begin{figure*}
    \includegraphics[width=0.8\textwidth]{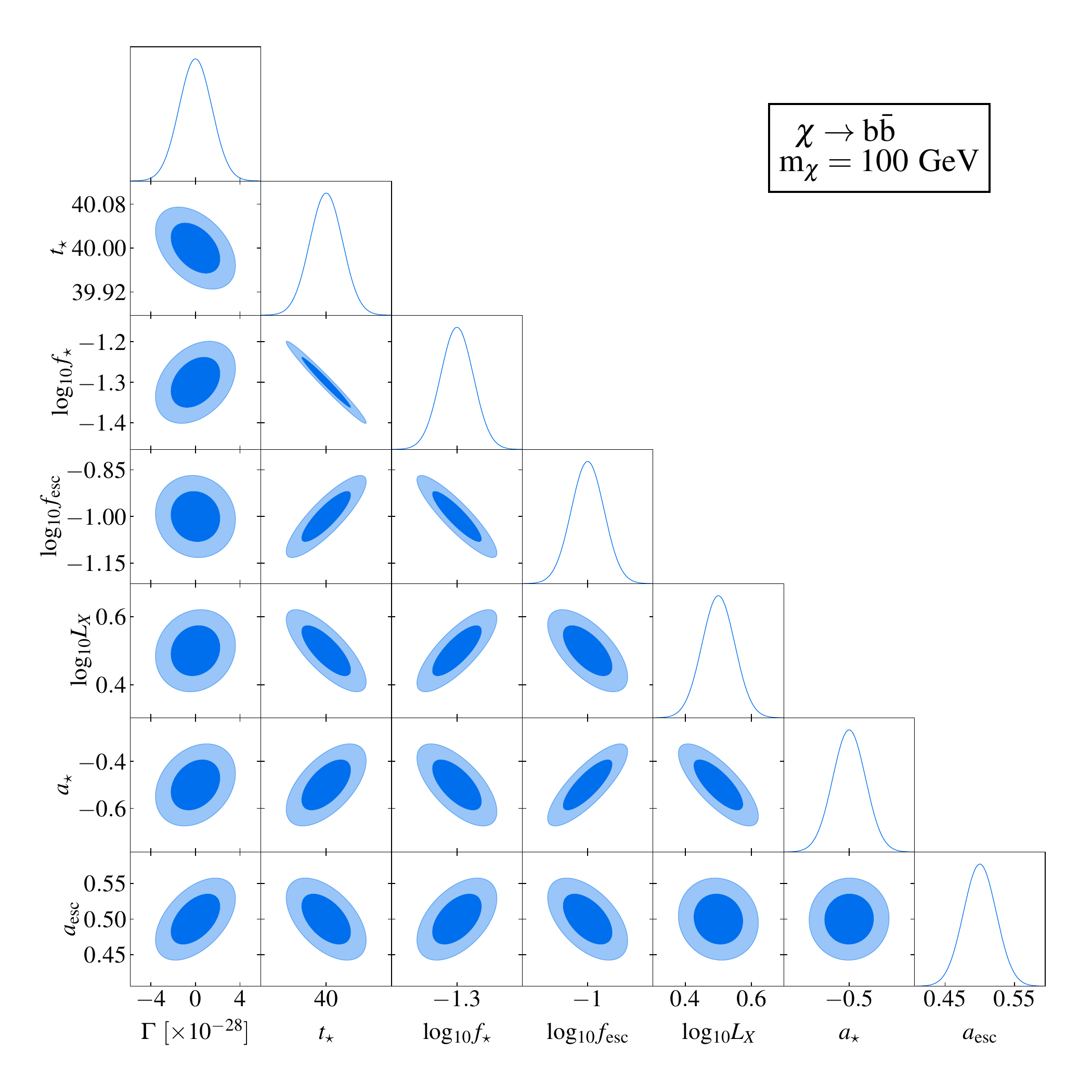}
    \caption{Same as Fig.~\ref{fig:ann_corry} but for $100$\,GeV \ac{DM} particles decaying into bottom-anti-bottom quark pairs.}
    \label{fig:dec_corrb}
\end{figure*}

Remaining results of Fisher matrix forecasts are shown in Figs.~\ref{fig:ann_corry}--\ref{fig:dec_corrb}.
Figs.~\ref{fig:ann_corry} and \ref{fig:ann_corrb} present the Fisher information matrix analysis results of \ac{DM} particle annihilation through channels producing photon pairs and bottom-anti-bottom quark pairs.  
The dark and light shaded regions correspond to $1\sigma$ and $2\sigma$ confidence level contours, respectively.  
The solid curves represent the marginalized posterior distributions of relevant parameters.  
The Fiducial parameter values are consistent with those in Fig.~\ref{fig:measurement error power}.  
For annihilation via the photon-pair channel, the \ac{DM} particle mass is 10\,MeV.
For the bottom-anti-bottom quark pair channel, it is 10\,GeV.  
Similarly, Figs.~\ref{fig:dec_corry} and \ref{fig:dec_corrb} present the Fisher matrix analysis results for DM particle decay through the same channels.  
The configuration settings are identical to those used in Figs.~\ref{fig:ann_corry} and \ref{fig:ann_corrb}.

\label{sec:appendix}

\acknowledgments

We are grateful to Kazunori Kohri, Yichao Li, and Xukun Zhang for their valuable contributions to the discussion. This work is supported by the National SKA Program of China (Grant Nos. 2022SKA0110200, 2022SKA0110203), the National Key R\&D Program of China (Grant No. 2023YFC2206403), the National Natural Science Foundation of China (Grant Nos. 12175243, 12473001, and 12533001), the China Manned Space
Program (Grant No. CMS-CSST-2025-A02), and the National 111 Project (Grant No. B16009).  

\newpage


\bibliographystyle{JHEP}
\bibliography{revtex.bib}

\providecommand{\href}[2]{#2}\begingroup\raggedright\begin{thebibliography}{10}

\bibitem{ParticleDataGroup:2024cfk}
{\scshape Particle Data Group} collaboration, \emph{{Review of particle
  physics}}, \href{https://doi.org/10.1103/PhysRevD.110.030001}{\emph{Phys.
  Rev. D} {\bfseries 110} (2024) 030001}.

\bibitem{Rubakov:2019lyf}
V.A.~Rubakov, \emph{{Cosmology and dark matter}},
  \href{https://doi.org/10.23730/CYRSP-2021-005.129}{\emph{CERN Yellow Rep.
  School Proc.} {\bfseries 5} (2022) 129}
  [\href{https://arxiv.org/abs/1912.04727}{{\ttfamily 1912.04727}}].

\bibitem{Bertone:2016nfn}
G.~Bertone and D.~Hooper, \emph{{History of dark matter}},
  \href{https://doi.org/10.1103/RevModPhys.90.045002}{\emph{Rev. Mod. Phys.}
  {\bfseries 90} (2018) 045002}
  [\href{https://arxiv.org/abs/1605.04909}{{\ttfamily 1605.04909}}].

\bibitem{Liu:2019zez}
H.~Liu, \emph{{Dark Matter Energy Deposition and Production from the Table-Top
  to the Cosmos}}, Ph.D. thesis, MIT, 2019.
\newblock \href{https://arxiv.org/abs/1907.04324}{{\ttfamily 1907.04324}}.

\bibitem{Chen:2003gz}
X.-L.~Chen and M.~Kamionkowski, \emph{{Particle decays during the cosmic dark
  ages}}, \href{https://doi.org/10.1103/PhysRevD.70.043502}{\emph{Phys. Rev. D}
  {\bfseries 70} (2004) 043502}
  [\href{https://arxiv.org/abs/astro-ph/0310473}{{\ttfamily
  astro-ph/0310473}}].

\bibitem{Thorpe-Morgan:2024zcq}
C.R.~Thorpe-Morgan, \emph{{Astrophysical Frontiers: The Indirect Detection of
  Dark Matter {\&} the Analysis of Gamma-ray Binary Systems}}, Ph.D. thesis,
  Universit{\"a}t T{\"u}bingen, Tubingen U., 10, 2024.
\newblock 10.15496/publikation-99791.

\bibitem{Gaskins:2016cha}
J.M.~Gaskins, \emph{{A review of indirect searches for particle dark matter}},
  \href{https://doi.org/10.1080/00107514.2016.1175160}{\emph{Contemp. Phys.}
  {\bfseries 57} (2016) 496}
  [\href{https://arxiv.org/abs/1604.00014}{{\ttfamily 1604.00014}}].

\bibitem{Bertone:2004pz}
G.~Bertone, D.~Hooper and J.~Silk, \emph{{Particle dark matter: Evidence,
  candidates and constraints}},
  \href{https://doi.org/10.1016/j.physrep.2004.08.031}{\emph{Phys. Rept.}
  {\bfseries 405} (2005) 279}
  [\href{https://arxiv.org/abs/hep-ph/0404175}{{\ttfamily hep-ph/0404175}}].

\bibitem{Hawking:1971ei}
S.~Hawking, \emph{{Gravitationally collapsed objects of very low mass}},
  \href{https://doi.org/10.1093/mnras/152.1.75}{\emph{Mon. Not. Roy. Astron.
  Soc.} {\bfseries 152} (1971) 75}.

\bibitem{Wang:2025lti}
X.~Wang, M.~Sasaki and Y.-l.~Zhang, \emph{{The Dual Primordial Black Hole
  Formation Scenario}},  \href{https://arxiv.org/abs/2505.09337}{{\ttfamily
  2505.09337}}.

\bibitem{Barroso:2024cgg}
E.J.~Barroso, L.F.~Dem{\'e}trio, S.D.P.~Vitenti and X.~Ye, \emph{{Primordial
  black hole formation in a dust bouncing model}},
  \href{https://doi.org/10.1088/1475-7516/2025/01/052}{\emph{JCAP} {\bfseries
  01} (2025) 052} [\href{https://arxiv.org/abs/2405.00207}{{\ttfamily
  2405.00207}}].

\bibitem{Carr:2021bzv}
B.~Carr and F.~Kuhnel, \emph{{Primordial black holes as dark matter
  candidates}},
  \href{https://doi.org/10.21468/SciPostPhysLectNotes.48}{\emph{SciPost Phys.
  Lect. Notes} {\bfseries 48} (2022) 1}
  [\href{https://arxiv.org/abs/2110.02821}{{\ttfamily 2110.02821}}].

\bibitem{Carr:2009jm}
B.J.~Carr, K.~Kohri, Y.~Sendouda and J.~Yokoyama, \emph{{New cosmological
  constraints on primordial black holes}},
  \href{https://doi.org/10.1103/PhysRevD.81.104019}{\emph{Phys. Rev. D}
  {\bfseries 81} (2010) 104019}
  [\href{https://arxiv.org/abs/0912.5297}{{\ttfamily 0912.5297}}].

\bibitem{Zhang:2023tfv}
C.~Zhang and X.~Zhang, \emph{{Gravitational capture of magnetic monopoles by
  primordial black holes in the early universe}},
  \href{https://doi.org/10.1007/JHEP10(2023)037}{\emph{JHEP} {\bfseries 10}
  (2023) 037} [\href{https://arxiv.org/abs/2302.07002}{{\ttfamily
  2302.07002}}].

\bibitem{Zhang:2023zmb}
C.~Zhang and X.~Zhang, \emph{{Magnetic monopole meets primordial black hole: an
  extended analysis}},
  \href{https://doi.org/10.1140/epjc/s10052-024-12383-8}{\emph{Eur. Phys. J. C}
  {\bfseries 84} (2024) 100}
  [\href{https://arxiv.org/abs/2308.07166}{{\ttfamily 2308.07166}}].

\bibitem{Jia:2025vqn}
N.~Jia, S.-S.~Bao, C.~Zhang, H.~Zhang and X.~Zhang, \emph{{Superradiant dark
  matter production from primordial black holes: Impact of multiple modes and
  gravitational wave emission}},
  \href{https://arxiv.org/abs/2504.18935}{{\ttfamily 2504.18935}}.

\bibitem{Mohapatra:2025qpz}
V.~Mohapatra, \emph{{Cosmological bounds on dark matter annihilation using dark
  ages 21-cm signal}},  \href{https://arxiv.org/abs/2506.20648}{{\ttfamily
  2506.20648}}.

\bibitem{Shao:2024owi}
Y.~Shao, T.-Y.~Sun, M.-L.~Zhao and X.~Zhang, \emph{{Analytical modeling of the
  one-dimensional power spectrum of 21-cm forest based on a halo model
  method}},  \href{https://arxiv.org/abs/2411.17094}{{\ttfamily 2411.17094}}.

\bibitem{Sun:2024ywb}
T.-Y.~Sun, Y.~Shao, Y.~Li, Y.~Xu, H.~Wang and X.~Zhang, \emph{{Deep
  learning-driven likelihood-free parameter inference for 21-cm forest
  observations}},
  \href{https://doi.org/10.1038/s42005-025-02139-5}{\emph{Commun. Phys.}
  {\bfseries 8} (2025) 220} [\href{https://arxiv.org/abs/2407.14298}{{\ttfamily
  2407.14298}}].

\bibitem{Nishizawa:2024bnh}
A.J.~Nishizawa, P.K.~Natwariya and K.~Kadota, \emph{{Machine learning
  constraints on dark matter annihilation during the epoch of reionization: A
  convolutional neural network analysis of the 21-cm signal}},
  \href{https://doi.org/10.1103/PhysRevD.111.083546}{\emph{Phys. Rev. D}
  {\bfseries 111} (2025) 083546}
  [\href{https://arxiv.org/abs/2410.04755}{{\ttfamily 2410.04755}}].

\bibitem{Zhao:2024jad}
M.-L.~Zhao, S.~Wang and X.~Zhang, \emph{{Prospects for probing dark matter
  particles and primordial black holes with the Hongmeng mission using the 21
  cm global spectrum at cosmic dawn}},
  \href{https://doi.org/10.1088/1475-7516/2025/07/039}{\emph{JCAP} {\bfseries
  07} (2025) 039} [\href{https://arxiv.org/abs/2412.19257}{{\ttfamily
  2412.19257}}].

\bibitem{Shao:2023agv}
Y.~Shao, Y.~Xu, Y.~Wang, W.~Yang, R.~Li, X.~Zhang et~al., \emph{{The 21-cm
  forest as a simultaneous probe of dark matter and cosmic heating history}},
  \href{https://doi.org/10.1038/s41550-023-02024-7}{\emph{Nature Astron.}
  {\bfseries 7} (2023) 1116}
  [\href{https://arxiv.org/abs/2307.04130}{{\ttfamily 2307.04130}}].

\bibitem{Hiroshima:2021bxn}
N.~Hiroshima, K.~Kohri, T.~Sekiguchi and R.~Takahashi, \emph{{Impacts of new
  small-scale N-body simulations on dark matter annihilations constrained from
  cosmological 21-cm line observations}},
  \href{https://doi.org/10.1103/PhysRevD.104.083547}{\emph{Phys. Rev. D}
  {\bfseries 104} (2021) 083547}
  [\href{https://arxiv.org/abs/2103.14810}{{\ttfamily 2103.14810}}].

\bibitem{Saha:2021pqf}
A.K.~Saha and R.~Laha, \emph{{Sensitivities on nonspinning and spinning
  primordial black hole dark matter with global 21-cm troughs}},
  \href{https://doi.org/10.1103/PhysRevD.105.103026}{\emph{Phys. Rev. D}
  {\bfseries 105} (2022) 103026}
  [\href{https://arxiv.org/abs/2112.10794}{{\ttfamily 2112.10794}}].

\bibitem{Cang:2021owu}
J.~Cang, Y.~Gao and Y.-Z.~Ma, \emph{{21-cm constraints on spinning primordial
  black holes}},
  \href{https://doi.org/10.1088/1475-7516/2022/03/012}{\emph{JCAP} {\bfseries
  03} (2022) 012} [\href{https://arxiv.org/abs/2108.13256}{{\ttfamily
  2108.13256}}].

\bibitem{Mena:2019nhm}
O.~Mena, S.~Palomares-Ruiz, P.~Villanueva-Domingo and S.J.~Witte,
  \emph{{Constraining the primordial black hole abundance with 21-cm
  cosmology}}, \href{https://doi.org/10.1103/PhysRevD.100.043540}{\emph{Phys.
  Rev. D} {\bfseries 100} (2019) 043540}
  [\href{https://arxiv.org/abs/1906.07735}{{\ttfamily 1906.07735}}].

\bibitem{Lopez-Honorez:2016sur}
L.~Lopez-Honorez, O.~Mena, {\'A}.~Molin{\'e}, S.~Palomares-Ruiz and
  A.C.~Vincent, \emph{{The 21 cm signal and the interplay between dark matter
  annihilations and astrophysical processes}},
  \href{https://doi.org/10.1088/1475-7516/2016/08/004}{\emph{JCAP} {\bfseries
  08} (2016) 004} [\href{https://arxiv.org/abs/1603.06795}{{\ttfamily
  1603.06795}}].

\bibitem{Xu:2024vdn}
C.~Xu, W.~Qin and T.R.~Slatyer, \emph{{CMB limits on decaying dark matter
  beyond the ionization threshold}},
  \href{https://doi.org/10.1103/PhysRevD.110.123529}{\emph{Phys. Rev. D}
  {\bfseries 110} (2024) 123529}
  [\href{https://arxiv.org/abs/2408.13305}{{\ttfamily 2408.13305}}].

\bibitem{Capozzi:2023xie}
F.~Capozzi, R.Z.~Ferreira, L.~Lopez-Honorez and O.~Mena, \emph{{CMB and
  Lyman-\ensuremath{\alpha} constraints on dark matter decays to photons}},
  \href{https://doi.org/10.1088/1475-7516/2023/06/060}{\emph{JCAP} {\bfseries
  06} (2023) 060} [\href{https://arxiv.org/abs/2303.07426}{{\ttfamily
  2303.07426}}].

\bibitem{Zhang:2023usm}
Z.-X.~Zhang, Y.-M.~Wang, J.~Cang, Z.~Zhang, Y.~Liu, S.-Y.~Li et~al.,
  \emph{{Dark matter search with CMB: a~study of foregrounds}},
  \href{https://doi.org/10.1088/1475-7516/2023/10/002}{\emph{JCAP} {\bfseries
  10} (2023) 002} [\href{https://arxiv.org/abs/2304.07793}{{\ttfamily
  2304.07793}}].

\bibitem{Acharya:2020jbv}
S.K.~Acharya and R.~Khatri, \emph{{CMB and BBN constraints on evaporating
  primordial black holes revisited}},
  \href{https://doi.org/10.1088/1475-7516/2020/06/018}{\emph{JCAP} {\bfseries
  06} (2020) 018} [\href{https://arxiv.org/abs/2002.00898}{{\ttfamily
  2002.00898}}].

\bibitem{Chluba:2020oip}
J.~Chluba, A.~Ravenni and S.K.~Acharya, \emph{{Thermalization of large energy
  release in the early Universe}},
  \href{https://doi.org/10.1093/mnras/staa2131}{\emph{Mon. Not. Roy. Astron.
  Soc.} {\bfseries 498} (2020) 959}
  [\href{https://arxiv.org/abs/2005.11325}{{\ttfamily 2005.11325}}].

\bibitem{Planck:2018vyg}
{\scshape Planck} collaboration, \emph{{Planck 2018 results. VI. Cosmological
  parameters}},
  \href{https://doi.org/10.1051/0004-6361/201833910}{\emph{Astron. Astrophys.}
  {\bfseries 641} (2020) A6}
  [\href{https://arxiv.org/abs/1807.06209}{{\ttfamily 1807.06209}}].

\bibitem{Clark:2016nst}
S.~Clark, B.~Dutta, Y.~Gao, L.E.~Strigari and S.~Watson, \emph{{Planck
  Constraint on Relic Primordial Black Holes}},
  \href{https://doi.org/10.1103/PhysRevD.95.083006}{\emph{Phys. Rev. D}
  {\bfseries 95} (2017) 083006}
  [\href{https://arxiv.org/abs/1612.07738}{{\ttfamily 1612.07738}}].

\bibitem{Kawasaki:2015yya}
M.~Kawasaki, K.~Kohri, T.~Moroi and Y.~Takaesu, \emph{{Revisiting Big-Bang
  Nucleosynthesis Constraints on Dark-Matter Annihilation}},
  \href{https://doi.org/10.1016/j.physletb.2015.10.048}{\emph{Phys. Lett. B}
  {\bfseries 751} (2015) 246}
  [\href{https://arxiv.org/abs/1509.03665}{{\ttfamily 1509.03665}}].

\bibitem{Lopez-Honorez:2013cua}
L.~Lopez-Honorez, O.~Mena, S.~Palomares-Ruiz and A.C.~Vincent,
  \emph{{Constraints on dark matter annihilation from CMB observationsbefore
  Planck}}, \href{https://doi.org/10.1088/1475-7516/2013/07/046}{\emph{JCAP}
  {\bfseries 07} (2013) 046} [\href{https://arxiv.org/abs/1303.5094}{{\ttfamily
  1303.5094}}].

\bibitem{Slatyer:2009yq}
T.R.~Slatyer, N.~Padmanabhan and D.P.~Finkbeiner, \emph{{CMB Constraints on
  WIMP Annihilation: Energy Absorption During the Recombination Epoch}},
  \href{https://doi.org/10.1103/PhysRevD.80.043526}{\emph{Phys. Rev. D}
  {\bfseries 80} (2009) 043526}
  [\href{https://arxiv.org/abs/0906.1197}{{\ttfamily 0906.1197}}].

\bibitem{LHAASO:2025wgl}
{\scshape LHAASO} collaboration, \emph{{All-sky search for individual
  Primordial Black Hole bursts with LHAASO}},
  \href{https://arxiv.org/abs/2505.24586}{{\ttfamily 2505.24586}}.

\bibitem{Saha:2024ies}
A.K.~Saha, A.~Singh, P.~Parashari and R.~Laha, \emph{{Hunting Primordial Black
  Hole Dark Matter in Lyman-$\alpha$ Forest}},
  \href{https://arxiv.org/abs/2409.10617}{{\ttfamily 2409.10617}}.

\bibitem{Yang:2024vij}
C.~Yang, S.~Wang, M.-L.~Zhao and X.~Zhang, \emph{{Search for the Hawking
  radiation of primordial black holes: prospective sensitivity of LHAASO}},
  \href{https://doi.org/10.1088/1475-7516/2024/10/083}{\emph{JCAP} {\bfseries
  10} (2024) 083} [\href{https://arxiv.org/abs/2408.10897}{{\ttfamily
  2408.10897}}].

\bibitem{Koechler:2023ual}
J.~Koechler, \emph{{X-rays constraints on sub-GeV Dark Matter}},  in \emph{{TeV
  Particle Astrophysics 2023}}, 9, 2023
  [\href{https://arxiv.org/abs/2309.10043}{{\ttfamily 2309.10043}}].

\bibitem{Calore:2022pks}
F.~Calore, A.~Dekker, P.D.~Serpico and T.~Siegert, \emph{{Constraints on light
  decaying dark matter candidates from 16~yr of INTEGRAL/SPI observations}},
  \href{https://doi.org/10.1093/mnras/stad457}{\emph{Mon. Not. Roy. Astron.
  Soc.} {\bfseries 520} (2023) 4167}
  [\href{https://arxiv.org/abs/2209.06299}{{\ttfamily 2209.06299}}].

\bibitem{Foster:2022nva}
J.W.~Foster, Y.~Park, B.R.~Safdi, Y.~Soreq and W.L.~Xu, \emph{{Search for dark
  matter lines at the Galactic Center with 14~years of Fermi data}},
  \href{https://doi.org/10.1103/PhysRevD.107.103047}{\emph{Phys. Rev. D}
  {\bfseries 107} (2023) 103047}
  [\href{https://arxiv.org/abs/2212.07435}{{\ttfamily 2212.07435}}].

\bibitem{Bernal:2022swt}
N.~Bernal, V.~Mu{\~n}oz-Albornoz, S.~Palomares-Ruiz and P.~Villanueva-Domingo,
  \emph{{Current and future neutrino limits on the abundance of primordial
  black holes}},
  \href{https://doi.org/10.1088/1475-7516/2022/10/068}{\emph{JCAP} {\bfseries
  10} (2022) 068} [\href{https://arxiv.org/abs/2203.14979}{{\ttfamily
  2203.14979}}].

\bibitem{Cirelli:2020bpc}
M.~Cirelli, N.~Fornengo, B.J.~Kavanagh and E.~Pinetti, \emph{{Integral X-ray
  constraints on sub-GeV Dark Matter}},
  \href{https://doi.org/10.1103/PhysRevD.103.063022}{\emph{Phys. Rev. D}
  {\bfseries 103} (2021) 063022}
  [\href{https://arxiv.org/abs/2007.11493}{{\ttfamily 2007.11493}}].

\bibitem{Wang:2020uvi}
S.~Wang, D.-M.~Xia, X.~Zhang, S.~Zhou and Z.~Chang, \emph{{Constraining
  primordial black holes as dark matter at JUNO}},
  \href{https://doi.org/10.1103/PhysRevD.103.043010}{\emph{Phys. Rev. D}
  {\bfseries 103} (2021) 043010}
  [\href{https://arxiv.org/abs/2010.16053}{{\ttfamily 2010.16053}}].

\bibitem{Boudaud:2018hqb}
M.~Boudaud and M.~Cirelli, \emph{{Voyager 1 $e^\pm$ Further Constrain
  Primordial Black Holes as Dark Matter}},
  \href{https://doi.org/10.1103/PhysRevLett.122.041104}{\emph{Phys. Rev. Lett.}
  {\bfseries 122} (2019) 041104}
  [\href{https://arxiv.org/abs/1807.03075}{{\ttfamily 1807.03075}}].

\bibitem{Boudaud:2018oya}
M.~Boudaud, T.~Lacroix, M.~Stref and J.~Lavalle, \emph{{Robust cosmic-ray
  constraints on $p$-wave annihilating MeV dark matter}},
  \href{https://doi.org/10.1103/PhysRevD.99.061302}{\emph{Phys. Rev. D}
  {\bfseries 99} (2019) 061302}
  [\href{https://arxiv.org/abs/1810.01680}{{\ttfamily 1810.01680}}].

\bibitem{HESS:2018kom}
{\scshape HESS} collaboration, \emph{{Searches for gamma-ray lines and 'pure
  WIMP' spectra from Dark Matter annihilations in dwarf galaxies with
  H.E.S.S}}, \href{https://doi.org/10.1088/1475-7516/2018/11/037}{\emph{JCAP}
  {\bfseries 11} (2018) 037}
  [\href{https://arxiv.org/abs/1810.00995}{{\ttfamily 1810.00995}}].

\bibitem{MAGIC:2017avy}
{\scshape MAGIC} collaboration, \emph{{Indirect dark matter searches in the
  dwarf satellite galaxy Ursa Major II with the MAGIC Telescopes}},
  \href{https://doi.org/10.1088/1475-7516/2018/03/009}{\emph{JCAP} {\bfseries
  03} (2018) 009} [\href{https://arxiv.org/abs/1712.03095}{{\ttfamily
  1712.03095}}].

\bibitem{VERITAS:2017tif}
{\scshape VERITAS} collaboration, \emph{{Dark Matter Constraints from a Joint
  Analysis of Dwarf Spheroidal Galaxy Observations with VERITAS}},
  \href{https://doi.org/10.1103/PhysRevD.95.082001}{\emph{Phys. Rev. D}
  {\bfseries 95} (2017) 082001}
  [\href{https://arxiv.org/abs/1703.04937}{{\ttfamily 1703.04937}}].

\bibitem{Boudaud:2016mos}
M.~Boudaud, J.~Lavalle and P.~Salati, \emph{{Novel cosmic-ray electron and
  positron constraints on MeV dark matter particles}},
  \href{https://doi.org/10.1103/PhysRevLett.119.021103}{\emph{Phys. Rev. Lett.}
  {\bfseries 119} (2017) 021103}
  [\href{https://arxiv.org/abs/1612.07698}{{\ttfamily 1612.07698}}].

\bibitem{Cohen:2016uyg}
T.~Cohen, K.~Murase, N.L.~Rodd, B.R.~Safdi and Y.~Soreq,
  \emph{{\ensuremath{\gamma} -ray Constraints on Decaying Dark Matter and
  Implications for IceCube}},
  \href{https://doi.org/10.1103/PhysRevLett.119.021102}{\emph{Phys. Rev. Lett.}
  {\bfseries 119} (2017) 021102}
  [\href{https://arxiv.org/abs/1612.05638}{{\ttfamily 1612.05638}}].

\bibitem{Carr:2016hva}
B.J.~Carr, K.~Kohri, Y.~Sendouda and J.~Yokoyama, \emph{{Constraints on
  primordial black holes from the Galactic gamma-ray background}},
  \href{https://doi.org/10.1103/PhysRevD.94.044029}{\emph{Phys. Rev. D}
  {\bfseries 94} (2016) 044029}
  [\href{https://arxiv.org/abs/1604.05349}{{\ttfamily 1604.05349}}].

\bibitem{Fermi-LAT:2015att}
{\scshape Fermi-LAT} collaboration, \emph{{Searching for Dark Matter
  Annihilation from Milky Way Dwarf Spheroidal Galaxies with Six Years of Fermi
  Large Area Telescope Data}},
  \href{https://doi.org/10.1103/PhysRevLett.115.231301}{\emph{Phys. Rev. Lett.}
  {\bfseries 115} (2015) 231301}
  [\href{https://arxiv.org/abs/1503.02641}{{\ttfamily 1503.02641}}].

\bibitem{Massari:2015xea}
A.~Massari, E.~Izaguirre, R.~Essig, A.~Albert, E.~Bloom and
  G.A.~G\'omez-Vargas, \emph{{Strong Optimized Conservative $Fermi$-LAT
  Constraints on Dark Matter Models from the Inclusive Photon Spectrum}},
  \href{https://doi.org/10.1103/PhysRevD.91.083539}{\emph{Phys. Rev. D}
  {\bfseries 91} (2015) 083539}
  [\href{https://arxiv.org/abs/1503.07169}{{\ttfamily 1503.07169}}].

\bibitem{HESS:2014zqa}
{\scshape H.E.S.S.} collaboration, \emph{{Search for dark matter annihilation
  signatures in H.E.S.S. observations of Dwarf Spheroidal Galaxies}},
  \href{https://doi.org/10.1103/PhysRevD.90.112012}{\emph{Phys. Rev. D}
  {\bfseries 90} (2014) 112012}
  [\href{https://arxiv.org/abs/1410.2589}{{\ttfamily 1410.2589}}].

\bibitem{Aleksic:2013xea}
J.~Aleksi\'c et~al., \emph{{Optimized dark matter searches in deep observations
  of Segue 1 with MAGIC}},
  \href{https://doi.org/10.1088/1475-7516/2014/02/008}{\emph{JCAP} {\bfseries
  02} (2014) 008} [\href{https://arxiv.org/abs/1312.1535}{{\ttfamily
  1312.1535}}].

\bibitem{Diamanti:2013bia}
R.~Diamanti, L.~Lopez-Honorez, O.~Mena, S.~Palomares-Ruiz and A.C.~Vincent,
  \emph{{Constraining Dark Matter Late-Time Energy Injection: Decays and P-Wave
  Annihilations}},
  \href{https://doi.org/10.1088/1475-7516/2014/02/017}{\emph{JCAP} {\bfseries
  02} (2014) 017} [\href{https://arxiv.org/abs/1308.2578}{{\ttfamily
  1308.2578}}].

\bibitem{Essig:2013goa}
R.~Essig, E.~Kuflik, S.D.~McDermott, T.~Volansky and K.M.~Zurek,
  \emph{{Constraining Light Dark Matter with Diffuse X-Ray and Gamma-Ray
  Observations}}, \href{https://doi.org/10.1007/JHEP11(2013)193}{\emph{JHEP}
  {\bfseries 11} (2013) 193} [\href{https://arxiv.org/abs/1309.4091}{{\ttfamily
  1309.4091}}].

\bibitem{Cirelli:2009bb}
M.~Cirelli, F.~Iocco and P.~Panci, \emph{{Constraints on Dark Matter
  annihilations from reionization and heating of the intergalactic gas}},
  \href{https://doi.org/10.1088/1475-7516/2009/10/009}{\emph{JCAP} {\bfseries
  10} (2009) 009} [\href{https://arxiv.org/abs/0907.0719}{{\ttfamily
  0907.0719}}].

\bibitem{Kolopanis:2022mgk}
M.~Kolopanis, J.C.~Pober, D.C.~Jacobs and S.~McGraw, \emph{{New EoR power
  spectrum limits from MWA Phase II using the delay spectrum method and novel
  systematic rejection}},
  \href{https://doi.org/10.1093/mnras/stad845}{\emph{Mon. Not. Roy. Astron.
  Soc.} {\bfseries 521} (2023) 5120}
  [\href{https://arxiv.org/abs/2210.10885}{{\ttfamily 2210.10885}}].

\bibitem{Trott:2020szf}
C.M.~Trott et~al., \emph{{Deep multiredshift limits on Epoch of Reionization 21
  cm power spectra from four seasons of Murchison Widefield Array
  observations}}, \href{https://doi.org/10.1093/mnras/staa414}{\emph{Mon. Not.
  Roy. Astron. Soc.} {\bfseries 493} (2020) 4711}
  [\href{https://arxiv.org/abs/2002.02575}{{\ttfamily 2002.02575}}].

\bibitem{Mertens:2025pvk}
F.G.~Mertens et~al., \emph{{Deeper multi-redshift upper limits on the Epoch of
  Reionization 21-cm signal power spectrum from LOFAR between z=8.3 and
  z=10.1}},  \href{https://arxiv.org/abs/2503.05576}{{\ttfamily 2503.05576}}.

\bibitem{2019MNRAS.488.4271G}
B.K.~{Gehlot}, F.G.~{Mertens}, L.V.E.~{Koopmans}, M.A.~{Brentjens},
  S.~{Zaroubi}, B.~{Ciardi} et~al., \emph{{The first power spectrum limit on
  the 21-cm signal of neutral hydrogen during the Cosmic Dawn at z = 20-25 from
  LOFAR}}, \href{https://doi.org/10.1093/mnras/stz1937}{\emph{mnras} {\bfseries
  488} (2019) 4271} [\href{https://arxiv.org/abs/1809.06661}{{\ttfamily
  1809.06661}}].

\bibitem{2022ApJ...925..221A}
Z.~{Abdurashidova}, J.E.~{Aguirre}, P.~{Alexander}, Z.S.~{Ali}, Y.~{Balfour},
  A.P.~{Beardsley} et~al., \emph{{First Results from HERA Phase I: Upper Limits
  on the Epoch of Reionization 21 cm Power Spectrum}},
  \href{https://doi.org/10.3847/1538-4357/ac1c78}{\emph{apj} {\bfseries 925}
  (2022) 221} [\href{https://arxiv.org/abs/2108.02263}{{\ttfamily
  2108.02263}}].

\bibitem{Gagnon-Hartman:2025oxd}
S.~Gagnon-Hartman, J.~Davies and A.~Mesinger, \emph{{Detecting galaxy-21-cm
  cross-correlation during reionization}},
  \href{https://arxiv.org/abs/2502.20447}{{\ttfamily 2502.20447}}.

\bibitem{Bierlich:2022pfr}
C.~Bierlich et~al., \emph{{A comprehensive guide to the physics and usage of
  PYTHIA 8.3}},
  \href{https://doi.org/10.21468/SciPostPhysCodeb.8}{\emph{SciPost Phys.
  Codeb.} {\bfseries 2022} (2022) 8}
  [\href{https://arxiv.org/abs/2203.11601}{{\ttfamily 2203.11601}}].

\bibitem{Cirelli:2010xx}
M.~Cirelli, G.~Corcella, A.~Hektor, G.~Hutsi, M.~Kadastik, P.~Panci et~al.,
  \emph{{PPPC 4 DM ID: A Poor Particle Physicist Cookbook for Dark Matter
  Indirect Detection}},
  \href{https://doi.org/10.1088/1475-7516/2012/10/E01}{\emph{JCAP} {\bfseries
  03} (2011) 051} [\href{https://arxiv.org/abs/1012.4515}{{\ttfamily
  1012.4515}}].

\bibitem{Liu:2019bbm}
H.~Liu, G.W.~Ridgway and T.R.~Slatyer, \emph{{Code package for calculating
  modified cosmic ionization and thermal histories with dark matter and other
  exotic energy injections}},
  \href{https://doi.org/10.1103/PhysRevD.101.023530}{\emph{Phys. Rev. D}
  {\bfseries 101} (2020) 023530}
  [\href{https://arxiv.org/abs/1904.09296}{{\ttfamily 1904.09296}}].

\bibitem{Slatyer:2015kla}
T.R.~Slatyer, \emph{{Indirect Dark Matter Signatures in the Cosmic Dark Ages
  II. Ionization, Heating and Photon Production from Arbitrary Energy
  Injections}}, \href{https://doi.org/10.1103/PhysRevD.93.023521}{\emph{Phys.
  Rev. D} {\bfseries 93} (2016) 023521}
  [\href{https://arxiv.org/abs/1506.03812}{{\ttfamily 1506.03812}}].

\bibitem{Slatyer:2015jla}
T.R.~Slatyer, \emph{{Indirect dark matter signatures in the cosmic dark ages.
  I. Generalizing the bound on s-wave dark matter annihilation from Planck
  results}}, \href{https://doi.org/10.1103/PhysRevD.93.023527}{\emph{Phys. Rev.
  D} {\bfseries 93} (2016) 023527}
  [\href{https://arxiv.org/abs/1506.03811}{{\ttfamily 1506.03811}}].

\bibitem{Slatyer:2012yq}
T.R.~Slatyer, \emph{{Energy Injection And Absorption In The Cosmic Dark Ages}},
  \href{https://doi.org/10.1103/PhysRevD.87.123513}{\emph{Phys. Rev. D}
  {\bfseries 87} (2013) 123513}
  [\href{https://arxiv.org/abs/1211.0283}{{\ttfamily 1211.0283}}].

\bibitem{Auffinger:2020ztk}
J.~Auffinger and A.~Arbey, \emph{{BlackHawk: A tool for computing Black Hole
  evaporation}}, \href{https://doi.org/10.22323/1.392.0024}{\emph{PoS}
  {\bfseries TOOLS2020} (2021) 024}
  [\href{https://arxiv.org/abs/2012.12902}{{\ttfamily 2012.12902}}].

\bibitem{Facchinetti:2023slb}
G.~Facchinetti, L.~Lopez-Honorez, Y.~Qin and A.~Mesinger, \emph{{21cm signal
  sensitivity to dark matter decay}},
  \href{https://doi.org/10.1088/1475-7516/2024/01/005}{\emph{JCAP} {\bfseries
  01} (2024) 005} [\href{https://arxiv.org/abs/2308.16656}{{\ttfamily
  2308.16656}}].

\bibitem{Pritchard:2011xb}
J.R.~Pritchard and A.~Loeb, \emph{{21-cm cosmology}},
  \href{https://doi.org/10.1088/0034-4885/75/8/086901}{\emph{Rept. Prog. Phys.}
  {\bfseries 75} (2012) 086901}
  [\href{https://arxiv.org/abs/1109.6012}{{\ttfamily 1109.6012}}].

\bibitem{Furlanetto:2006jb}
S.~Furlanetto, S.P.~Oh and F.~Briggs, \emph{{Cosmology at Low Frequencies: The
  21 cm Transition and the High-Redshift Universe}},
  \href{https://doi.org/10.1016/j.physrep.2006.08.002}{\emph{Phys. Rept.}
  {\bfseries 433} (2006) 181}
  [\href{https://arxiv.org/abs/astro-ph/0608032}{{\ttfamily
  astro-ph/0608032}}].

\bibitem{Mesinger:2010ne}
A.~Mesinger, S.~Furlanetto and R.~Cen, \emph{{21cmFAST: A Fast, Semi-Numerical
  Simulation of the High-Redshift 21-cm Signal}},
  \href{https://doi.org/10.1111/j.1365-2966.2010.17731.x}{\emph{Mon. Not. Roy.
  Astron. Soc.} {\bfseries 411} (2011) 955}
  [\href{https://arxiv.org/abs/1003.3878}{{\ttfamily 1003.3878}}].

\bibitem{Mason:2022obt}
C.A.~Mason, J.B.~Mu\~noz, B.~Greig, A.~Mesinger and J.~Park, \emph{{21cmfish:
  Fisher-matrix framework for fast parameter forecasts from the cosmic 21-cm
  signal}}, \href{https://doi.org/10.1093/mnras/stad2145}{\emph{Mon. Not. Roy.
  Astron. Soc.} {\bfseries 524} (2023) 4711}
  [\href{https://arxiv.org/abs/2212.09797}{{\ttfamily 2212.09797}}].

\bibitem{Pober:2012zz}
J.C.~Pober, A.R.~Parsons, D.R.~DeBoer, P.~McDonald, M.~McQuinn, J.E.~Aguirre
  et~al., \emph{{The Baryon Acoustic Oscillation Broadband and Broad-beam
  Array: Design Overview and Sensitivity Forecasts}},
  \href{https://doi.org/10.1088/0004-6256/145/3/65}{\emph{Astron. J.}
  {\bfseries 145} (2013) 65} [\href{https://arxiv.org/abs/1210.2413}{{\ttfamily
  1210.2413}}].

\bibitem{Zahn:2010yw}
O.~Zahn, A.~Mesinger, M.~McQuinn, H.~Trac, R.~Cen and L.E.~Hernquist,
  \emph{{Comparison Of Reionization Models: Radiative Transfer Simulations And
  Approximate, Semi-Numeric Models}},
  \href{https://doi.org/10.1111/j.1365-2966.2011.18439.x}{\emph{Mon. Not. Roy.
  Astron. Soc.} {\bfseries 414} (2011) 727}
  [\href{https://arxiv.org/abs/1003.3455}{{\ttfamily 1003.3455}}].

\bibitem{Braun:2019gdo}
R.~Braun, A.~Bonaldi, T.~Bourke, E.~Keane and J.~Wagg, \emph{{Anticipated
  Performance of the Square Kilometre Array -- Phase 1 (SKA1)}},
  \href{https://arxiv.org/abs/1912.12699}{{\ttfamily 1912.12699}}.

\bibitem{frechet1943extension}
M.~Fr{\'e}chet, \emph{Sur l'extension de certaines {\'e}valuations statistiques
  au cas de petits {\'e}chantillons}, {\emph{Revue de l'Institut International
  de Statistique} (1943) 182}.

\bibitem{Mesinger:2013nua}
A.~Mesinger, A.~Ewall-Wice and J.~Hewitt, \emph{{Reionization and beyond:
  detecting the peaks of the cosmological 21 cm signal}},
  \href{https://doi.org/10.1093/mnras/stu125}{\emph{Mon. Not. Roy. Astron.
  Soc.} {\bfseries 439} (2014) 3262}
  [\href{https://arxiv.org/abs/1310.0465}{{\ttfamily 1310.0465}}].

\bibitem{Chhabra:2025dfo}
M.~Chhabra and S.~Bharadwaj, \emph{{Probing the HI distribution at small scales
  using 21-cm Intensity Mapping at large scales}},
  \href{https://arxiv.org/abs/2508.19126}{{\ttfamily 2508.19126}}.

\bibitem{Gill:2025ydz}
S.S.~Gill, S.~Bharadwaj, K.M.A.~Elahi, S.K.~Sethi and A.K.~Patwa, \emph{{The
  EoR 21-cm Bispectrum at $z=8.2$ from MWA data I: Foregrounds and preliminary
  upper limits}},  \href{https://arxiv.org/abs/2507.04964}{{\ttfamily
  2507.04964}}.

\bibitem{Noble:2024uzl}
L.~Noble, M.~Kamran, S.~Majumdar, C.S.~Murmu, R.~Ghara, G.~Mellema et~al.,
  \emph{{Impact of the Epoch of Reionization sources on the 21-cm bispectrum}},
  \href{https://doi.org/10.1088/1475-7516/2024/10/003}{\emph{JCAP} {\bfseries
  10} (2024) 003} [\href{https://arxiv.org/abs/2406.03118}{{\ttfamily
  2406.03118}}].

\bibitem{Diao:2025adp}
K.~Diao and Y.~Mao, \emph{{Multi-fidelity emulator for large-scale 21 cm
  lightcone images: a few-shot transfer learning approach with generative
  adversarial network}},  \href{https://arxiv.org/abs/2502.04246}{{\ttfamily
  2502.04246}}.

\bibitem{Shimabukuro:2025equ}
H.~Shimabukuro, Y.~Xu and Y.~Shao, \emph{{Analyzing the 21cm forest with
  Wavelet Scattering Transform: Insight into non-Gaussian features of the 21cm
  forest}},  \href{https://arxiv.org/abs/2504.14656}{{\ttfamily 2504.14656}}.

\bibitem{Zhao:2023tep}
X.~Zhao, Y.~Mao, S.~Zuo and B.D.~Wandelt, \emph{{Simulation-based Inference of
  Reionization Parameters from 3D Tomographic 21 cm Light-cone Images. II.
  Application of Solid Harmonic Wavelet Scattering Transform}},
  \href{https://doi.org/10.3847/1538-4357/ad5ff0}{\emph{Astrophys. J.}
  {\bfseries 973} (2024) 41}
  [\href{https://arxiv.org/abs/2310.17602}{{\ttfamily 2310.17602}}].

\bibitem{Diao:2024wrf}
K.~Diao, Z.~Chen, X.~Chen and Y.~Mao, \emph{{Reionization Parameter Inference
  from 3D Minkowski Functionals of the 21 cm Signals}},
  \href{https://doi.org/10.3847/1538-4357/ad6c40}{\emph{Astrophys. J.}
  {\bfseries 974} (2024) 141}
  [\href{https://arxiv.org/abs/2406.20058}{{\ttfamily 2406.20058}}].

\bibitem{2025arXiv250612129L}
Z.~{Li}, Z.~{Cai}, X.~{Wang}, Z.~{Li}, A.~{Dekel}, K.C.~{Sarkar} et~al.,
  \emph{{A 13-Billion-Year View of Galaxy Growth: Metallicity Gradient
  Evolution from the Local Universe to $z=9$ with JWST and Archival Surveys}},
  \href{https://doi.org/10.48550/arXiv.2506.12129}{\emph{arXiv e-prints} (2025)
  arXiv:2506.12129} [\href{https://arxiv.org/abs/2506.12129}{{\ttfamily
  2506.12129}}].

\end{thebibliography}\endgroup

\end{document}